\documentclass[twocol]{ametsocV5}

\usepackage{amsmath,amsfonts,amssymb,bm}
\usepackage{mathptmx}
\usepackage{newtxtext}
\usepackage{newtxmath}
\usepackage{multirow}
\usepackage{multicol, blindtext}

\title{Reforecasting two heavy-precipitation events with three convection-permitting ensembles}

\authors{Valerio Capecchi\correspondingauthor{Valerio Capecchi, capecchi@lamma.toscana.it}}

\affiliation{LaMMA, Laboratorio di Meteorologia e Modellistica Ambientale 
             per lo sviluppo sostenibile, Firenze, Italia}

\abstract{We investigate the potential added value of running three limited-area ensemble systems 
(with the WRF, Meso-NH and MOLOCH models and a grid spacing of approximately 2.5 km) for two heavy-precipitation 
events in Italy. Such high-resolution ensembles include an explicit treatment of convective processes and dynamically 
downscale the ECMWF global data, which have a grid spacing of approximately 18 km. 
The predictions are verified against rain-gauge data and their accuracy is evaluated over that of the 
driving coarser-resolution ensemble system. Furthermore, we compare the simulation speed 
(defined as the ratio of simulation length to wall-clock time) of the three limited-area models 
to estimate the computational effort for operational convection-permitting ensemble 
forecasting. We also study how the simulation wall-clock time scales with increasing numbers of computing 
elements (from 36 to 1152 cores). Objective verification methods generally show that convection-permitting 
forecasts outperform global forecasts for both events, although precipitation peaks remain 
largely underestimated for one of the two events.
Comparing simulation speeds, the MOLOCH model is the fastest and the Meso-NH is the slowest one. 
The WRF model attains efficient scalability, whereas it is limited for the Meso-NH and MOLOCH models 
when using more than 288 cores.
We finally demonstrate how the model simulation speed has the largest impact on a joint evaluation with the model performance because the accuracy of the three limited-area ensembles, amplifying the forecasting capability of the global predictions, does not differ substantially.
}

\begin{document}

\maketitle

%
%
\statement
We reforecasted two heavy-precipitation events that struck Italy in autumn 2011 by 
using recent versions of three limited-area weather models, which use a more accurate 
description of convective processes than the driving global model. By reforecasting past 
events, the aim of this work is to assess the information content carried by current 
numerical forecasting systems in the event of similar high-impact events happening again. 
In view of the potential use for operational forecasting, this study evaluates how the 
time needed to deliver the forecasts decreases as the computer speed increases. It is 
shown that such a scaling capability depends not only on the computing elements but 
also on the geometry of the model domain.
%


%
\section{Introduction}\label{sec:intro}
To account for the chaotic nature of weather systems, international centers introduced 
ensemble prediction systems (EPS) more than 25 years ago \citep{toth1993ensemble,tracton1993operational,buizza1995singular,molteni1996ecmwf}. 
Today, technological capability allows for running convection-permitting (CP) model 
ensembles with horizontal mesh sizes less than 3 km. To give a few examples, CP 
ensembles are operational in  national weather facilities in Europe 
\citep{gebhardt2008experimental,montani2011seven,raynaud2015comparison,hagelin2017met,klasa2018evaluation,frogner2019harmoneps}, 
the United States \citep{schwartz2015ncar,schwartz2015real} and Asia \citep{duc2013spatial}. 
The design of such high-resolution ensemble systems is complex and implies the trade-off between 
computational resources and the limited knowledge of the factors leading to error growth and 
propagation \citep{hohenegger2007predictability}. High-resolution ensembles were often found 
to be underdispersive, leading to overconfident predictions, in particular for high-impact 
weather \citep{vie2011cloud,romine2014representing}. Focusing on two severe 
weather events that occurred in Italy, in this study, we evaluate the reliability of three high-resolution ensembles 
using different CP models.

Following \cite{buizza2019introduction}, the design of an ensemble system has to account for (i) the initial perturbation strategy, (ii) the model uncertainty simulation strategy and (iii) the resolution, forecast length and number of members. \cite{schwartz2014characterizing} provided evidence that the main source of error for their CP ensemble system is uncertainty in the initial conditions. This has been confirmed by other works (see \citeauthor{wang2011central} \citeyear{wang2011central} for a review), and thus investigating techniques that better perturb initial small-scale features is a vital field of research. An ensemble of data assimilation (EDA) technique was demonstrated to provided added value to ensemble spread estimation, in particular when weak synoptic forcings are met \citep{vie2011cloud,bouttier2016sensitivity}. To contend with the high computational cost of such EDA systems, \cite{raynaud2015comparison} developed a less expensive method that randomly samples the model's background error covariance matrix. The authors obtained results as skillful as the more CPU-demanding EDA technique. \cite{leoncini2010perturbation} used a Monte Carlo approach to perturb initial conditions. The authors assessed the value of a simple stochastic technique to perturb the model state and studied how the ensemble members diverge from the control run as a consequence of the introduced uncertainties. Methods to address the model uncertainty comprise stochastically perturbed physics tendencies \citep{bouttier2012impact,romine2014representing}, stochastically perturbed parameterization \citep{fresnay2012heavy,jankov2019stochastically}, multiphysics \citep{berner2015increasing,loken2019spread,gasperoni2020comparison} or multimodel approaches \citep{ebert2001ability,eckel2005aspects,clark2019comparisons}. Some papers sought an optimal compromise between the model resolution and the ensemble size. Both \cite{schwartz2017toward} and \cite{raynaud2017impact} found that decreasing the grid spacing is beneficial for short-range forecasts, whereas increasing the number of members has positive impacts at longer lead times. It is also relevant to question the perturbations of the lateral boundary conditions \citep{leoncini2010perturbation,romine2014representing,bouttier2018clustering}. In fact, it has been proven that such perturbations control the ensemble spread from 12 hours of integration onward \citep{hohenegger2008cloud,peralta2012accounting}.

In the recent literature it has been suggested that the baseline approach of dynamical downscaling using CP models nested in a global ensemble with a coarser horizontal grid spacing provides valuable information. In fact, by capturing the large-scale flow that is the dominant driver of variability, it has been demonstrated that dynamical downscaling provides more realistic rainfall amounts than the driving system when strong synoptic forcings are present \citep{hohenegger2008cloud,kuhnlein2014impact}. However, compared to other CPU-intensive methods, dynamical downscaling suffers from a spin-up time of approximately 6-12 hours \citep{hohenegger2008cloud,vie2011cloud,kuhnlein2014impact,raynaud2015comparison}.

Since March 2016, the Integrated Forecasting System (IFS) model of the European Centre for Medium‐Range Weather Forecasts (ECMWF) has adopted the new cubic-octahedral grid, which replaced the linear reduced Gaussian grid employed previously. The spectral resolution of the ensemble forecasts was updated to TCO639 with a grid spacing of approximately 18 km \citep{malardel2016new}. The recent availability of such higher-resolution global ensemble data allows us to estimate the technical feasibility and value of the simple dynamical downscaling method to initialize a limited-area and CP model directly nested in the ECMWF global ensemble. This is a pragmatic approach to take full advantage of improved global data. It is foreseen that in the coming years, the horizontal resolution of global ensemble data will be further increased with the explicit goal of providing accurate predictions of high-impact weather in the medium-range \citep{roadmap2016}. This assumption suggests that small-scale uncertainties in CP ensembles might be more accurately addressed by sampling the larger-scale EPS.

This paper presents the results produced in the framework of two ECMWF Special Projects, the computational resources of which were granted during the years 2016-2018 and 2019-2021. The common goal of the two projects is to assess the added value of running a limited-area CP ensemble in terms of quantitative precipitation forecast (QPF). The accuracy of the cascade of state-of-the-art ensembles, from global to local, is evaluated by reforecasting past high-impact precipitation events and using three different mesoscale models. The dynamical downscaling method is chosen to start the regional ensembles, and the forecast lengths considered are longer than 24 hours. The regional weather models used are: the Weather Research and Forecasting (WRF) model \citep{skamarock2008description}, the Mesoscale Non-Hydrostatic (Meso-NH) model \citep{lac2018overview} and the MOdello LOCale in Hybrid coordinates (MOLOCH) model \citep{malguzzi20061966}. We present the ensemble reforecasts of two heavy precipitating events that affected the Liguria region (northwestern Italy) in autumn 2011: the flooding of the Cinque Terre site occurring on the 25th of October (hereinafter the CT case) and the flash flood of the city of Genoa occurring on the 4th of November (hereinafter the GE case). Both cases received attention from the scientific community because of the atmospheric processes involved and due to their dramatic impacts on the territory  \citep{rebora2013extreme,buzzi2014heavy,fiori2014analysis,capecchi2015statistical,cassola2015numerical,davolio2015effects,hally2015hydrometeorological,tiesi2016heavy}. The comparison of the results obtained with these three models contributes to the debate regarding their strengths and weaknesses with respect to (i) the accuracy of the results for the two events considered and (ii) the computational costs in view of the potential use for operational ensemble forecasting.

\section{The CT and GE cases}\label{sec:cases}
Although detailed descriptions of the CT and GE cases can be found in the literature, we include a short summary of both to make this paper self-contained.

\subsection{Descriptions of the precipitation events}\label{subsec:description}
\begin{figure}[t]
  \noindent\includegraphics[height=\columnwidth,angle=-90]{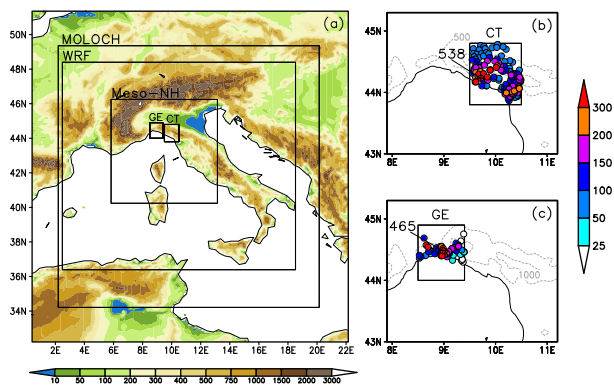}\\
  \caption{Panel (a): topography and geographical extent of the three domains of integration for the limited-area models. Panel (b): area of interest of the CT case with observed rainfall data registered by rain-gauges on the 25th of October 2011. Panel (c): area of interest of the GE case with observed rainfall data registered by rain-gauges on the 4th of November 2011. The maximum observed value is reported for each event (units of mm/day). In panels (b) and (c), the gray dashed contours indicate elevations of 500 and 1000 meters above sea level.}\label{fig:domain}
\end{figure}

Locations along the coast of Liguria near the Cinque Terre site and foothills of the Apennine Mountains in northern Tuscany (northwestern Italy, see Figure \ref{fig:domain}b) were affected by an extreme precipitation event on the 25th of October 2011 \citep{rebora2013extreme} that satisfied ingredients favoring severe weather over complex orography \citep{rotunno2001mechanisms,miglietta2009numerical}. The event occurred in association with rainfall rates exceeding 150 mm/1-hour, 450 mm/12-hour and 538 mm/24-hour at the rain-gauge of Brugnato-Borghetto Vara, resulting in several landslides and debris flows in the Magra and Vara basins \citep{capecchi2015statistical}.
A few days later, on the 4th of November, a storm affected the city of Genoa, located in the central part of the Liguria region (see Figure \ref{fig:domain}c). A self-regenerating and quasi-stationary V-shaped mesoscale convective system  originated a few kilometers away from the city and observed maximum rainfall rates reached approximately 390 mm/6-hour and 460 mm/24-hour \citep{rebora2013extreme}, causing Bisagno Creek to flood urban areas.

\cite{buzzi2014heavy} outlined the common synoptic-scale features between the CT and GE cases. Both events were characterized by upper-level troughs over the eastern Atlantic/western Mediterranean Sea and low-level ridges over the central/eastern Mediterranean Sea, which forced the troughs to advance eastward slowly. Strong vertical shear and diffluent flow in the mid-upper troposphere over Liguria favored the development of intense mesoscale convective systems in both cases. \cite{buzzi2014heavy} spotted the presence of two main mesoscale flows: (i) moist, warm southeasterly winds over the Ligurian Gulf impinging the Ligurian Apennines and (ii) a northerly outflow of cold air from the Po Valley (see Figure 4 in  \citeauthor{rebora2013extreme}, \citeyear{rebora2013extreme}). The convergence line between these two flows in combination with orographic uplift triggered convective precipitation over the area.
\begin{table*}[h]
\caption{Setup of the key characteristics of the limited-area and convection-permitting ensemble forecasts.}\label{tab:setup}
\centering
\begin{tabular}{p{0.1\linewidth}p{0.15\linewidth}p{0.15\linewidth}p{0.15\linewidth}p{0.15\linewidth}p{0.1\linewidth}}
\hline\hline
\textbf{Model} & \textbf{Grid Spacing (km)} & \textbf{Rows$\times$Columns} & \textbf{Vertical Levels} & \textbf{Grid Points} & \textbf{Time Step (sec)}
\\
&$\Delta x$ & $R\times C$& $L$& $R\times C\times L$  & $\Delta t$         
\\
\hline
WRF & 3& $400\times440$& 55 & $\simeq 9.7$ million & 18 
\\
Meso-NH& 2.5 & $225\times270$& 52 & $\simeq 3.1$ million & 6 
\\
MOLOCH &2.5 & $514\times614$   & 50 & $\simeq 15.4$ million & 30 
\\
\hline
\end{tabular}
\end{table*}
\subsection{Past numerical experiments}
The CT case was analyzed by \cite{buzzi2014heavy} and \cite{davolio2015effects} with the MOLOCH model and by \cite{capecchi2015statistical} and \cite{cassola2015numerical} with the WRF model. \cite{buzzi2014heavy} tested the model’s ability to predict rainfall by varying both the initial/boundary conditions and the grid spacing of the model. The best results were obtained with a 1.5 km grid spacing. Although the simulated precipitation maxima were underestimated, the authors found good agreement with the localization of rainfall patterns. This was achieved owing to the accurate prediction of the low-level convergence line between the warm and moist southeasterly flow over the Tyrrhenian Sea and the colder northerly air flow originating from the Ligurian Apennines and Po Valley. A similar study was conducted by \cite{davolio2015effects}, who tested the effects of increasing horizontal resolution of the MOLOCH model to improve the quality of the atmospheric forcings in a hydrological forecasting chain. They recommended a mesh size of less than 2 km to achieve a beneficial impact on hydrological predictions. From a meteorological point of view, the authors underlined how the air mass associated with the southerly/southeasterly low-level flow was more unstable for the CT case than for the GE case. \cite{capecchi2015statistical} simulated the CT case with the WRF model and integrated numerical weather forecasts with static geomorphological, geological and climatological information to feed a statistical model for the prediction of rainfall-induced shallow landslides. Although the grid spacing of their forecasts (3 km) was greater than the characteristic scale of landslide modeling (micro-$\gamma$ scale), they found a positive impact of QPFs on the landslide occurrence predictions. \cite{cassola2015numerical} conducted sensitivity tests by varying the microphysics scheme in the WRF model and found that the Thompson scheme \citep{thompson2008explicit} was among those that provided the best overall performing outputs.

\cite{fiori2014analysis} performed a hindcast of the GE case with the WRF model. They implemented two-way nested domains with mesh sizes of 5 and 1 km. They tested different cumulus parameterizations and microphysical schemes; the lengths of their experiments were 36 and 24 hours. The authors found that the model's performance is sensitive to both to the time of initialization (with poor results for the 36-hour length forecast) and the microphysical settings. By tuning the number of initial cloud droplets in the Thompson scheme, the authors gained improved results in regard to not only the value of peak rainfall but also the spatial pattern of precipitation. \cite{hally2015hydrometeorological} simulated the GE case using a CP multimodel ensemble (namely, WRF-NMM, WRF-ARW, Meso-NH and AROME) with 30 members. The model settings exhibited large differences regarding the extent of the domain, the number of vertical levels, the grid spacing  (ranging from 2.5 km to 0.5 km), the initial/boundary conditions, etc \citep{hally2015hydrometeorological}. To account for the model uncertainties, the authors introduced random perturbations on the physical processes of the Meso-NH model, following the method proposed in \cite{fresnay2012heavy}.
The Meso-NH model forced by French global model data gave the more realistic reconstruction of rainfall distribution among the deterministic runs, whereas the WRF-ARW simulation displayed the most intense rainfall value. According to the authors conclusions, only the multimodel approach was able to predict the various aspects of the flash flood and provided useful inputs for hydrological modeling. \cite{tiesi2016heavy} used the NOAA's Local Analysis and Prediction System to account for uncertainties in the initial condition and provide starting data for the MOLOCH model. They assimilated both conventional (surface and sounding) and unconventional (satellite and radar reflectivity) data and found that the assimilated run outperformed the control run in terms of QPF.

\section{Models and numerical setup}
In the following, we provide a short overview of the models used in this study. We stress again the fact that they all are set with the explicit treatment of convective processes.

The WRF model is the result of joint efforts by U.S. governmental agencies and the University of Oklahoma. It is a fully compressible, Eulerian, nonhydrostatic mesoscale model; we implemented the Advanced Research WRF (ARW) version of the model, which is described in \cite{skamarock2008description}. 
The microphysics option adopted is detailed in \cite{thompson2008explicit}; it is a single-moment parameterization and explicitly predicts the mixing ratios of five liquid and ice species: cloud water, rain, cloud ice, snow, and graupel.

Meso-NH is a French research community model, jointly developed by the Centre National des Recherches M\'et\'eorologiques and Laboratoire d'A\'erologie at the Universit\'e Paul Sabatier. It is designed to simulate the time evolution of several atmospheric variables ranging from the large meso-$\alpha$  scale ($\simeq$ 2000 $km$) down to the micro-$\gamma$ scale ($\simeq$ 20 $m$), typical of the large eddy simulations. For a general overview of the Meso-NH model and its applications, see \cite{lac2018overview}.
Regarding microphysics, we set the one-moment ICE3 scheme \citep{caniaux1994numerical}, taking five water species into account: cloud droplets, raindrops, pristine ice crystals, snow or aggregates, and graupel.

The MOLOCH model is a non-hydrostatic, fully compressible model that uses a hybrid terrain-following coordinate, relaxing smoothly to horizontal surfaces. It is developed at the Institute of Atmospheric Sciences and Climate (ISAC) of the Italian National Research Council (CNR). Details about the model can be found in \cite{malguzzi20061966}.
It was initially developed for research purposes, but today it is being used operationally by various regional meteorological services both in Italy and abroad. The microphysical scheme is based on the parameterization proposed in \cite{drofa2004parameterization}, which describes the interactions of cloud water, cloud ice, rain, snow and graupel.

In Table \ref{tab:setup}, we summarize a few basic settings of the integration domain for the three models, namely, the grid spacing (expressed in km), the number of rows, columns and vertical levels, the resulting total number of grid points and the time step (expressed in seconds). The grid spacing is set to 3 km for the WRF model and 2.5 km for the Meso-NH and MOLOCH models. The extent of the horizontal grid is not the same among the three models; the extents are shown in Figure \ref{fig:domain}. The number of vertical levels spans from 50 (for the MOLOCH model) to 55 (for the WRF model). With these settings we obtain a number of grid points, for the three dimensional grid, ranging from approximately 3.1 million for the Meso-NH model to approximately 9.7 million for the WRF model and up to approximately 15.4 million for the MOLOCH model. To satisfy the Courant-Friedrichs-Lewy (CFL) stability condition, we set the time step to 18, 6 and 30 seconds for the WRF, Meso-NH and MOLOCH models, respectively.

We use different compilers and compilation options to build the executables on the ECMWF supercomputer. Some details are given in the Appendix.

\section{Data and methods}\label{sec:data_methods}
In March 2016, ECMWF introduced significant improvements to all the components of IFS. Changes in the new cycle (labeled 41r2) involved: the high-resolution, the medium-range ensemble, and the ensemble of data assimilations. One of the major updates was the introduction of the new cubic-octahedral grid, which has a higher resolution and is globally more uniform than the previous grid. Additional information on the implementation of the new grid is available in \cite{malardel2016new}. In this study, ensemble reforecasts (hereinafter ENS) of the CT and GE cases were produced using the IFS model cycle 41r2 with a horizontal grid spacing of approximately 18 km. Perturbations to the initial conditions of ENS were generated by combining singular vectors and EDA-perturbations generated as described in \cite{isaksen2010ensemble}. The construction of the singular vector perturbations is described in \cite{leutbecher2008ensemble}, whereas the EDA-perturbations are combined with singular vectors following \cite{lang2015impact}. Starting dates that have been considered for the CT (GE) case are from 00 UTC 23 October (2 November) 2011 to 12 UTC 24 October (3 November) 2011, every 12 hours. The ending dates are 00 UTC 26 October 2011 for CT and 00 UTC 5 November 2011 for GE so that the forecast length ranges from 72 hours to 36 hours for both cases; forecast lengths shorter than 36 hours were not considered.  See Table \ref{tab:starting} for a summary of the simulations and the codes adopted to name each forecast. In the following text, we use the acronyms WRF-ENS, MNH-ENS and MOL-ENS to refer to the CP ensembles produced using the WRF, Meso-NH and MOLOCH models, respectively, and using the ENS data as the initial and boundary conditions. The number of members of each CP ensemble system is the same as that in the ENS data (i.e., 50 members).
\begin{table}[]
\caption{Summary of the numerical simulations performed. The third column indicates the codes adopted to name the forecasts.}\label{tab:starting}
\begin{center}
\begin{tabular}{ccccccccc}
\hline\hline
\textbf{CT case} & \textbf{Starting date} & \textbf{Forecast length to} & \textbf{Forecast}\\
\textbf{       } & \textbf{of the simulations} & \textbf{00 UTC 26 Oct 2011} & \textbf{code}\\
\hline
                 & 00 UTC 23 Oct 2011  & 72 hours & CT$+$72h\\
                 & 12 UTC 23 Oct 2011  & 60 hours & CT$+$60h\\
                 & 00 UTC 24 Oct 2011  & 48 hours & CT$+$48h\\
                 & 12 UTC 24 Oct 2011  & 36 hours & CT$+$36h\\
\hline\hline
\textbf{GE case} & \textbf{Starting date} & \textbf{Forecast length to} & \textbf{Forecast}\\
\textbf{       } & \textbf{of the simulations} & \textbf{00 UTC 5 Nov 2011} & \textbf{code}\\
\hline
                 & 00 UTC 2 Nov 2011  & 72 hours & GE$+$72h\\
                 & 12 UTC 2 Nov 2011  & 60 hours & GE$+$60h\\
                 & 00 UTC 3 Nov 2011  & 48 hours & GE$+$48h\\
                 & 12 UTC 3 Nov 2011  & 36 hours & GE$+$36h\\
\hline
\end{tabular}
\end{center}
\end{table}
The QPF data are compared with observed precipitation amounts collected at the rain-gauges belonging to the inset boxes shown in Figure \ref{fig:domain} (panels on the right). Such boxes are chosen subjectively by drawing a $1^\circ\times1^\circ$ square around the areas for which the rain-gauges registered the highest precipitation amounts. The total numbers of rain-gauges are 149 and 55 for the CT and GE cases, respectively. The basic statistics of the daily accumulated precipitation observed during the two events are reported in Table \ref{tab:obse}. When presenting or discussing the results regarding the verification of the model predictions against observed data, we refer to the precipitation that occurred in the 24-hour period ending at 00 UTC 26 October (5 November) 2011 for the CT (GE) case.
\begin{table}[]
\caption{Percentiles and maximum values of the daily rainfall data observed during the CT and GE cases.}\label{tab:obse}
\begin{center}
\begin{tabular}{ccccccccc}
\hline\hline
& \textbf{25th percentile} & \textbf{Median} & \textbf{75th percentile} & \textbf{Max}
\\
\hline
\textbf{CT} & 59 & 110 & 168 & 538\\
\textbf{GE} & 80 & 130 & 172 & 465\\
\hline
\end{tabular}
\end{center}
\end{table}
To reduce the effects of the double-penalty error \citep{ebert2009neighborhood}, when extracting the QPF values at rain-gauge locations, we picked the four nearest-neighbor grid values and averaged them to provide the forecast value at that location. The performance of the ensemble mean, chosen as the representative member of each ensemble system, is assessed by looking at the performance diagrams \citep{roebber2009visualizing}. Such diagrams plot four measures of the dichotomous forecast: probability of detection (POD), success ratio (SR), bias and critical success index (CSI). Using the $2\times2$ contingency table for the dichotomous (yes/no) forecast shown in Table \ref{tab:contingency}, the four skill measures are defined as follows:
\begin{align*} 
POD  = & \frac{A}{A+C},\\ 
SR   = & 1 - \frac{B}{A+B},\\
bias = & \frac{A+B}{A+C},\\
CSI  = & \frac{A}{A+B+C}.
\end{align*}
To estimate potential heavy rainfall, we evaluate the maps of the probability of precipitation (PoP) exceeding predefined thresholds. The PoP is a common ensemble-based product, which expresses the occurrence probability of an extreme event measured by the fraction of ensemble members that predict a value higher than a predefined threshold. The probabilistic skills of the CP ensembles are compared to those of ENS by constructing the receiver operating characteristic (ROC) curve and calculating the area under it \citep{mason1982model}. The ROC curve contrasts the hit rate versus false alarm rate, using a set of increasing probability thresholds to make the yes/no decision. The area under the ROC curve is frequently used as an index of accuracy of an ensemble system in order to be able to discriminate between the occurrence and nonoccurrence of weather events; the higher the value is, the better it is, with 1 as the upper limit and values below 0.5 indicate no skill compared to a random forecast.

Following the notations of \cite{coiffier2011fundamentals}, the simulation speed of the generic model $M$ is defined as the ratio between the forecast length $H$ over the time $T_M$ required to end the simulation. It can be expressed by the following relationship:
\begin{equation}\label{eq:sim_speed}
\frac{H}{T_M}=\frac{\Delta t_M  \cdot S}{N_{v,M} \cdot N_{c,M}}
\end{equation}
where $\Delta t_M$ is the time step, which depends on the grid spacing $\Delta x_M$ and has to satisfy the CFL condition. The numerator $S$ is a measure of the computational speed (e.g., the number of processing elements or the floating operations per second). The term $N_{v,M}$ is the number of variables to be processed at each time step $\Delta t_M$ and depends on the number of grid points (i.e., the number of rows, columns and vertical levels) of the integration domain. The term $N_{c,M}$ represents the number of calculations to be made at each time step $\Delta t_M$ and is a function of the computational cost required by the numerical method used to solve the equations. In view of a possible use for operational ensemble forecasting, we compare how the simulation speed, defined by the left-hand side of equation \ref{eq:sim_speed}, of model $M$ scales as the computer speed $S$ increases. The factor $\frac{H}{T_i}$ is taken as a measure of actual time-to-solution. In this study, the index $M$ can assume the value of WRF, Meso-NH, or MOLOCH. We stress the fact that this evaluation is not biased towards either the number of grid points of the integration domain or the time step adopted. In fact, with the settings summarized in Table \ref{tab:setup}, the ratio $N_{v,M}/\Delta t_M$ is almost constant for all the CP models.
\begin{table}[]
\caption{The $2\times2$ contingency table.}\label{tab:contingency}
\begin{center}
\begin{tabular}{l|lcc}
\hline\hline
\multirow{2}{*}{}                                   &     & \multicolumn{2}{c}{Event Observed} \\\hline
                                                    &     & yes              & no              \\
\multicolumn{1}{c|}{\multirow{2}{*}{Event Forecast}} & yes & A                & B               \\
\multicolumn{1}{c|}{}                                & no  & C                & D              \\\hline
\end{tabular}
\end{center}
\end{table}
To jointly evaluate a numerical weather model $M$ in terms of its simulation speed $S_M$ and performance $P_M$, we heuristically define the linear integrated speed-performance ($LISP$) index as a linear combination of $S_M$ and $P_M$, namely:
\begin{equation}\label{eq:lisp}
{LISP_M}(\alpha)=(1-\alpha)S_M + \alpha P_M,
\end{equation}
where the scalar $\alpha \in [0,1] \subseteq \mathbb{R}$ is a weight such that if $\alpha=1$, then the $LISP_M$ index is conditioned on having the more accurate forecast data, regardless of the time needed to accomplish the simulation, whereas if $\alpha=0$ then the $LISP_M$ index weights the faster forecast (provided that the accuracy satisfies some minimum requirement, i.e. ROC area $>0.5$). We note that, if we choose the ROC area as a measure of the performance $P_M$, we have that $LISP_M\ge 0, \forall \alpha\in  [0,1]$ and the higher the value is, the better it is. If we have to evaluate models $M$ and $N$ on the basis of both the performance and speed, we can evaluate if: 
\begin{displaymath}
LISP_M \ge LISP_N,
\end{displaymath}
that is if:
\begin{displaymath}
P_M \ge P_N +\kappa(S_N-S_M).
\end{displaymath}
where, for sake of simplicity, we set $\kappa=\left(\frac{1-\alpha}{\alpha}\right)$.

\section{Results}\label{sec:results}
\subsection{Model performance: precipitation verification}\label{sec:precip_verif}
\begin{figure}
  \noindent\includegraphics[height=\columnwidth,angle=-90]{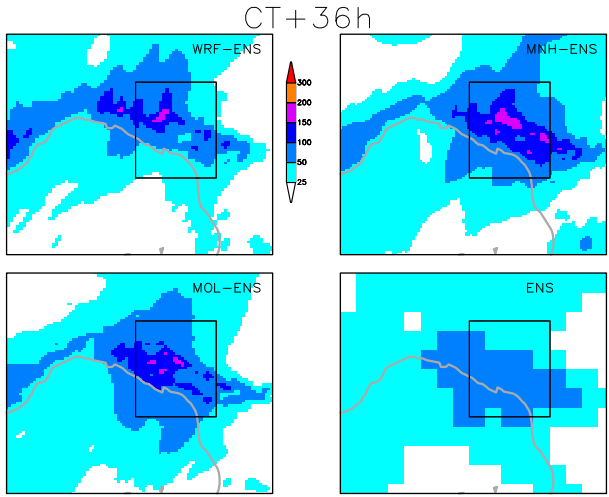}\\
  \caption{CT case: 24-hour accumulated precipitations for the WRF-ENS (top-left panel), MNH-ENS (top-right panel), MOL-ENS (bottom-left panel) and ENS (bottom-right panel) ensemble mean forecast. The forecast length is 36 hours.}\label{fig:ensmeanCT}
\end{figure}

\begin{figure}
  \noindent\includegraphics[height=\columnwidth,angle=-90]{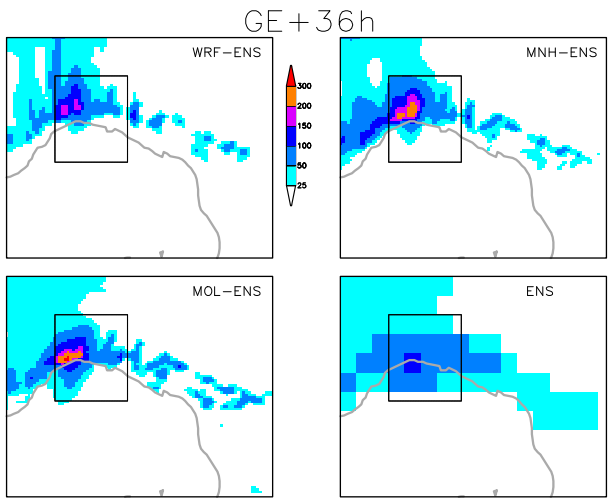}\\
  \caption{As in Figure \ref{fig:ensmeanCT} but for the  GE case.}\label{fig:ensmeanGE}
\end{figure}

\begin{figure}
  \noindent\includegraphics[height=\columnwidth,angle=-90]{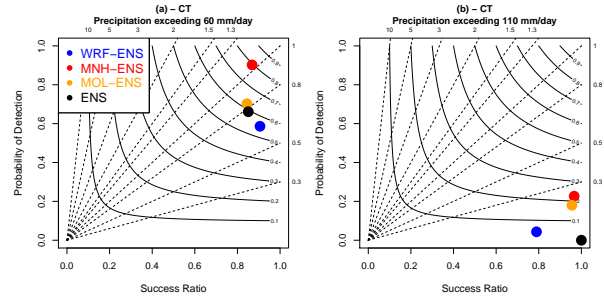}\\
  \caption{CT case: performance diagram of the ensemble mean forecast of WRF-ENS (blue), MNH-ENS (red), MOL-ENS (orange) and ENS (black). The X-axis shows the success ratio (SR), the Y-axis shows the probability of detection (POD), the curved lines represent the critical success index (CSI) values, and the dashed diagonal lines represent the bias. Panel (a) shows the scores for the precipitation threshold corresponding to the 25th percentile of the observed accumulated rainfall. Panel (b) as in (a) but for the 50th percentile.}\label{fig:PerfDiagCT}
\end{figure}

\begin{figure}
  \noindent\includegraphics[height=\columnwidth,angle=-90]{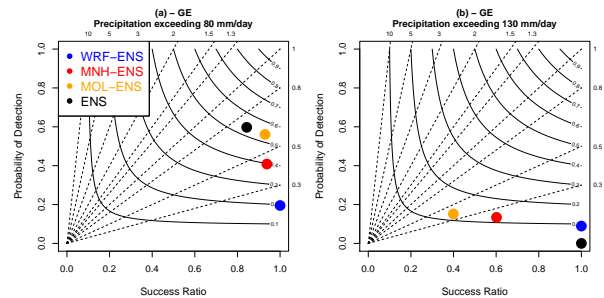}\\
  \caption{As in Figure \ref{fig:PerfDiagCT} but for the GE case.}\label{fig:PerfDiagGE}
\end{figure}

In Figures \ref{fig:ensmeanCT} and \ref{fig:ensmeanGE}, we show the QPF ensemble mean maps for CT+36h and GE+36h; longer-range forecasts do not provide substantial differences. The visual comparison of such maps with observed precipitation patterns shown in panels (b) and (c) of Figure \ref{fig:domain} suggests that the predicted ensemble means strongly underestimate the rainfall for both cases. To quantitatively analyze such underestimation, in Table \ref{tab:rmse},

\begin{table}
\caption{Root mean square error (RMSE) and mean error (ME) between the observed and predicted 24-hour accumulated precipitation values for the CT and GE cases.}\label{tab:rmse}
\begin{center}
\begin{tabular}{ccccccccc}
\hline\hline
& & \textbf{WRF-ENS} & \textbf{MNH-ENS} & \textbf{MOL-ENS} & \textbf{ENS}
\\
\hline
\textbf{CT} & RMSE (mm) & 103	& 89	& 93  & 102\\
            & ME   (mm) & -64	& -44	& -56 & -63\\
\textbf{GE} & RMSE (mm) & 138	& 129	& 126 & 123\\
            & ME   (mm) & -94	& -71	& -59 & -61\\
\hline
\end{tabular}
\end{center}
\end{table}
 we show the root mean square errors (RMSEs) and mean errors (MEs) between the predicted and observed precipitation values \citep{wilks2011statistical}; the values are averaged among all the forecast lengths. For the CT case, the RMSEs range from 89 mm (for the MNH-ENS ensemble mean) to 103 mm (for the WRF-ENS ensemble mean). The MNH-ENS ensemble mean also provides the best ME (i.e., closest to 0). For the GE case, we obtain an average RMSE of approximately 129 mm, with the ENS ensemble mean providing the lower value (123 mm) and WRF-ENS providing the higher value (138 mm). The analysis of the ME produces similar conclusions.
To further assess the accuracy of the ensemble means, in Figures \ref{fig:PerfDiagCT} and \ref{fig:PerfDiagGE},
we show the performance diagrams for the CT and GE cases, respectively. To make the yes/no decision, we chose two precipitation thresholds corresponding to the 25th percentile and the median of the observed precipitation data (see Table \ref{tab:obse}); the scores are averaged among all the forecast ranges. For the precipitation threshold equal to the 25th percentile (see panel (a) in the Figures), MNH-ENS provides more skillful predictions regarding CT; in fact, the score lies close to the top-right corner. For the GE case, ENS outperforms the other systems, having a bias of approximately 0.7 and relatively higher POD and CSI scores (approximately 0.6 for both). Considering the precipitation threshold equal to the median of the observed data (see panel (b) in the Figures), all the ensemble systems provide poor results (i.e. all the scores lie close to the bottom-right corner).

\begin{figure}[t]
  \noindent\includegraphics[height=\columnwidth,angle=-90]{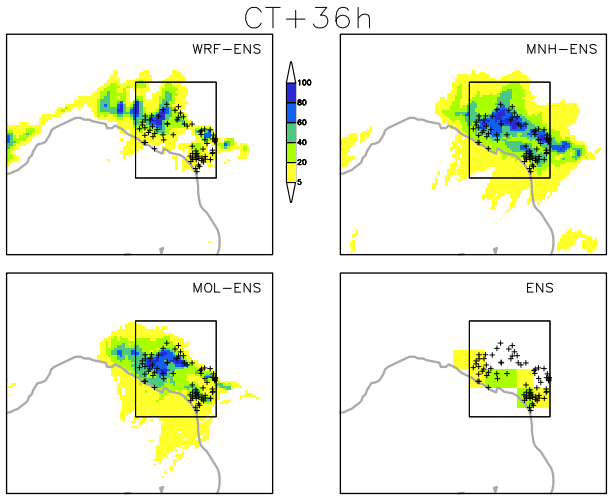}\\
  \caption{CT case: probability of precipitation (PoP) in excess of 110 mm (corresponding to approximately the median of the observed rainfall) for the 24-hour period ending on the 26th of October 2011 at 00 UTC. The forecast length is 36 hours.}\label{fig:PoPCTperc50}
\end{figure}

\begin{figure}[t]
  \noindent\includegraphics[height=\columnwidth,angle=-90]{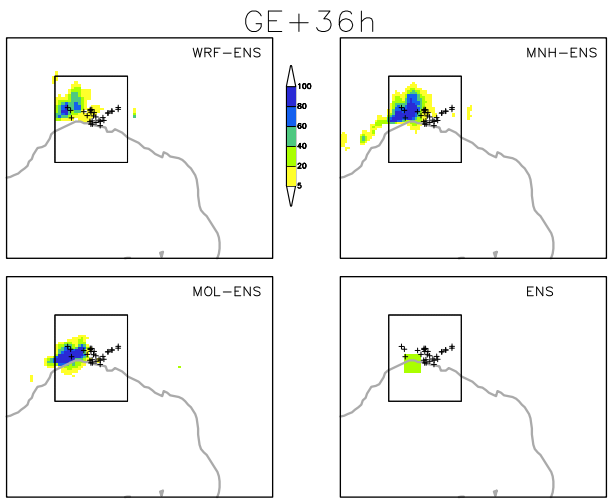}\\
  \caption{As in Figure \ref{fig:PoPCTperc50} but for the GE case. The precipitation thresholds corresponding to the median of the observed rainfall is approximately 130 mm.}\label{fig:PoPGEperc50}
\end{figure}

\begin{figure}[t]
  \noindent\includegraphics[width=\columnwidth,angle=0]{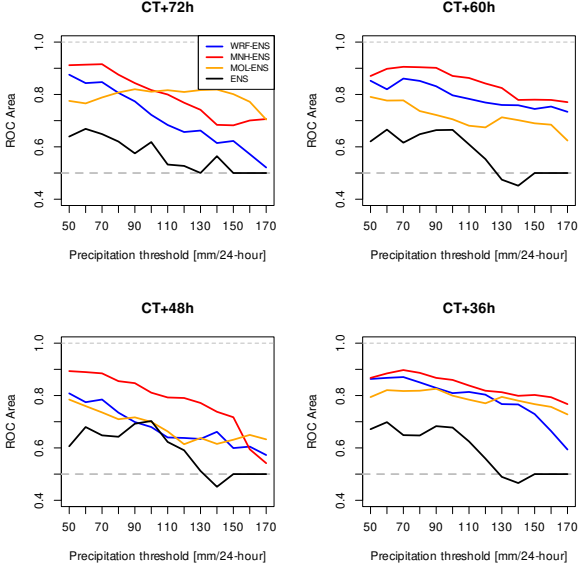}\\
  \caption{CT case: area under the receiver operating characteristic (ROC) curve as a function of the precipitation threshold for WRF-ENS (blue), MNH-ENS (red), MOL-ENS (orange) and ENS (black) data. The forecast lengths are  72 hours (top-left panel), 60 hours (top-right panel), 48 hours (bottom-left panel) and 36 hours (bottom-right panel).}\label{fig:ROCACT}
\end{figure}

\begin{figure}[t]
  \noindent\includegraphics[width=\columnwidth,angle=0]{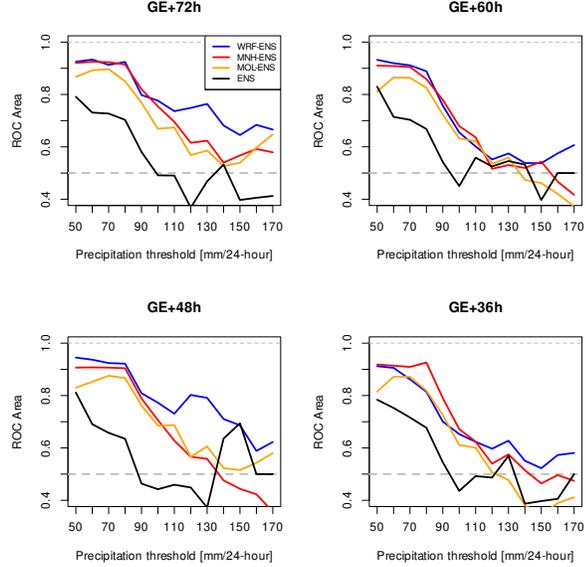}\\
  \caption{As in Figure \ref{fig:ROCACT} but for the GE case.}\label{fig:ROCAGE}
\end{figure}

Since the QPF ensemble means for both cases are skillful only for the precipitation threshold equal to the 25th percentile of the observed values, we also evaluated some ensemble-based verification metrics. In Figures \ref{fig:PoPCTperc50} and \ref{fig:PoPGEperc50}, we show the PoP maps for the four ensemble systems exceeding the medians of the observed values, which correspond to approximately 110 and 130 mm for the CT and GE cases, respectively. The forecast length is 36 hours, and longer-range forecasts do not provide results that differ substantially. Crosses represent the locations of rain-gauges where rainfall amounts greater than the threshold were actually registered. As regards the CT case (see Figure \ref{fig:PoPCTperc50}), the visual agreement between observations (Figure \ref{fig:domain} panel (b)) and PoP patterns appears good for all the ensembles. On average, the PoP values extracted at the rain-gauge locations are approximately 24\%, 51\%, 35\% and 10\% for WRF-ENS, MNH-ENS, MOL-ENS and ENS, respectively. Only ENS fails to produce valuable information (i.e., PoP values $<$5\%) for 27 out of 75 rain-gauges located in the northern part of the CT box (see the bottom-right panel in Figure \ref{fig:PoPCTperc50}). As regards the GE case, the PoP values are concentrated in a small portion of the domain and follow the pattern of the observations (see Figure \ref{fig:domain} panel (c)). MNH-ENS and MOL-ENS produce darker shaded areas than WRF-ENS, causing higher false alarm ratios. In fact, we found that 1, 5 and 4 out of 26 locations that did not record 130 mm of rainfall were located within the PoP$>$50\% contour for WRF-ENS, MNH-ENS and MOL-ENS, respectively. The ENS PoP map (bottom-right panel in Figure \ref{fig:PoPGEperc50}) has only one grid-point with a PoP value greater than 20\%, but the misplaced position of this point causes both the yes and the no events to be incorrectly predicted.
To quantitatively evaluate the probabilistic skills of the ensembles, we show the area underneath the ROC curve for the CT and GE cases in Figures \ref{fig:ROCACT} and \ref{fig:ROCAGE}, respectively, by varying the precipitation thresholds on the X-axis and considering different forecast ranges. The upper limit on the X-axis is set to the 75th percentile of the observed rainfall amount (which corresponds to approximately 170 mm for both cases). In general, the CP forecasts outperform the ENS forecasts (i.e., the CP curves lie above the ENS ones) for all thresholds and for all lead times with a few exceptions (e.g., CT+48h and GE+48h). As regards the CT case (Figure \ref{fig:ROCACT}), the ENS skill drops below the critical value of 0.5 approximately at the 130-mm precipitation threshold, whereas the CP ensembles provide valuable information (i.e., ROC area $>0.5$) up to 170 mm and beyond (plots not shown). Concerning the GE case (Figure \ref{fig:ROCAGE}), the profiles of the CP ensemble systems are very similar to each other; however, WRF-ENS data provide better results for GE+72h and GE+48h (i.e., a higher ROC area of approximately 0.78 on average for both forecasts). The CP ensemble (ENS) curves approach the 0.5 horizontal line when evaluating precipitation thresholds in the interval 130-140 mm (90-100 mm). A summary of the ROC area analysis is reported in Table \ref{tab:ROCA} in which the values are averaged among all the forecast ranges.
\begin{table}[t]
\caption{Areas under the ROC curve for the three convection-permitting forecasts and the ENS global predictions. For each  event, the values are averages among all the forecast lengths (from +72 to +36 hours) and among all the precipitation thresholds shown in Figures \ref{fig:ROCACT} and \ref{fig:ROCAGE}. For each event, the maximum value is highlighted in bold.}\label{tab:ROCA}
\begin{center}
\begin{tabular}{ccccccccc}
\hline\hline
& \textbf{WRF-ENS} & \textbf{MNH-ENS} & \textbf{MOL-ENS} & \textbf{ENS}
\\
\hline
\textbf{CT} & 0.742	& $\mathbf{0.815}$	& 0.744	& 0.580\\
\textbf{GE} & $\mathbf{0.739}$	& 0.683	& 0.653	& 0.558\\
\hline
\end{tabular}
\end{center}
\end{table}
To assess the capability of the ensembles to predict rainfall peaks close to the actually observed peaks, we extracted the maximum QPF value predicted by each member of each ensemble. In Figures \ref{fig:boxplotCTMAX} and \ref{fig:boxplotGEMAX}, we show the distributions (in the form of boxplots) of the QPF maxima for CT and GE, respectively. The boxplots demonstrated that ENS maxima are considerably lower than the CP maxima (approximately one-half for CT and one-third for GE). We also note that for GE, members of the CP ensembles provide QPF maxima close to or higher than the maximum observed values (indicated with the dashed horizontal line). For the CT case, none of the members provide QPF values close to the observed peak ($\simeq$ 538 mm).

To investigate the physical mechanisms underlying the CT and GE cases, in Figure \ref{fig:MNHconv}, we show the 3-hour accumulated rainfall and the 10-meter wind speed and direction (averaged over the same time period) of a single member of MNH-ENS for CT+36h (panels on the left) and GE+36h (panels on the right). The black point indicates the location of the rain-gauge that registered the maximum rainfall rate. In both cases, a convergence line is visible over the Ligurian Sea and marks the initiation of convective rainfall \citep{buzzi2014heavy}. In the CT case, the precipitation band oscillates from the east (panel (a)) to the west (panels (b) and (c)), and the resulting rainfall pattern is widespread over the whole area of interest. In the GE case, the position of the convergence line is steady, and thus, the precipitation pattern is limited to a small portion of the Genoa area.

\subsection{Model scaling}\label{sec:model_perf}
In light of the potential use for operational forecasting, in Figure \ref{fig:runtime} we show the scalability of the simulation speed, defined as the ratio of simulated time to elapsed wall-clock time, by varying (namely by doubling at each step) the number of cores used to realize a 36-hour long simulation (the CT+36h forecast). The values shown on the Y-axis are obtained by averaging the simulation speeds of five selected members, taken as representative of the speed of the whole ensemble system. The wall-clock time taken into account, considers only the period spent to compute the evolution of the state variables and not that spent for reading the initial conditions and postprocessing the model outputs. The MOLOCH model turns out to be the fastest, being on average approximately 2.3 times faster than the WRF model and approximately 5.3 times faster than the Meso-NH model.
To visualize the improvement in speed performance, in Figure \ref{fig:acc} we show the reduction, in percentage, of the wall-clock time when doubling the number of cores. The three models exhibit a fairly satisfactory reduction (i.e., $\simeq$ 50\%) up to 288 cores, and then the gain degrades rapidly as the number of cores increases, dropping below 25\% when using 1152 cores. Figure \ref{fig:acc} also shows the number of horizontal grid points assigned to each core (say $N_{x,y}$) with the colored labels. In fact, domains are decomposed into horizontal patches, and each computing element (namely, core) is responsible for a single patch. The Meso-NH model is poorly sensitive to the patch size through $\simeq$ 200 $N_{x,y}$, whereas the MOLOCH model is strongly limited by the patch size when $N_{x,y}$ is less than $\simeq $1000. The WRF model has a good scaling up to $\simeq$ 600 $N_{x,y}$, and then the elapsed wall-clock time is further reduced by approximately 35\% when $N_{x,y}$ is $\simeq$ 300.

\subsection{Model performance vs scaling}\label{sec:model_vs}
In Figure \ref{fig:lisp}, we show an analysis based on the ${LISP}$ index defined in equation \ref{eq:lisp}. Panel (a) refers to the CT case and panel (b) to GE. As proxy data for the performance of the forecasts ($P_M$ in equation \ref{eq:lisp}), we selected the ROC area values shown in Figures \ref{fig:ROCACT} and \ref{fig:ROCAGE}. Panel (a) takes into account the average ROC values of the four panels in Figure \ref{fig:ROCACT}; panel (b) as in panel (a) but averaging the data across the four panels in Figure \ref{fig:ROCAGE}. As a measure of the simulation speed ($S_M$ in equation \ref{eq:lisp}), we selected the simulation speed of the CP systems when running the CT+36h forecast with 288 cores (see Figure \ref{fig:runtime}). Both $S_M$ and $P_M$ values were normalized to constrain them in the interval $[0,1]\subseteq \mathbb{R}$. In panel (a) of Figure \ref{fig:lisp}, we evaluate, varying the  precipitation thresholds on the X-axis, the more accurate forecasts (as regards the CT case) against the fastest ones, namely we show $P_{MNH-ENS}$ (red line) and $P_{MOL-ENS}+\kappa (S_{MOL-ENS}-S_{MNH-ENS})$ (orange line). Panel (b) as in panel (a) but looking at the GE case. We set $\kappa=1/9$, that is we give more importance to the performance of the ensemble than to its speed (90\% vs. 10\%). Panel (a) shows that, for precipitation threshold higher (lower) than 120 mm, looking at the MOL-ENS (MNH-ENS) forecasts is more reliable considering both the performance and the speed of the ensemble. From panel (b) we can appreciate how looking at the slower WRF-ENS predictions represents a better trade-off between performance and speed than looking at the faster MOL-ENS predictions, for all the precipitation thresholds greater or equal to 120 mm.
\begin{figure}[t]
  \noindent\includegraphics[width=\columnwidth,angle=0]{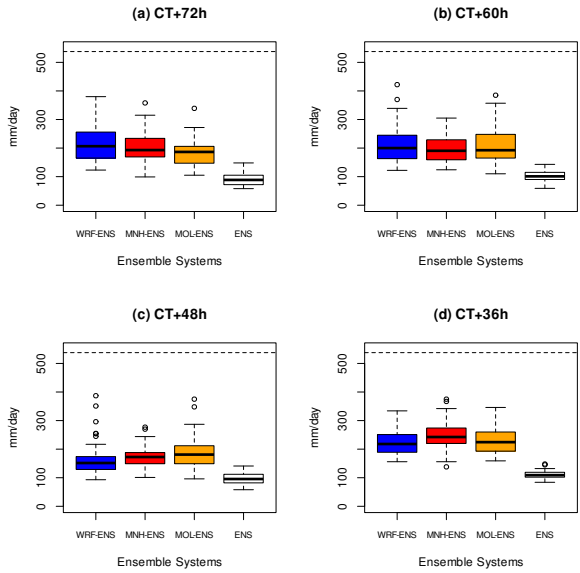}\\
  \caption{CT case: boxplot of the QPF maxima provided by each EPS in the area of interest depicted in Figure \ref{fig:domain}. The  lower and upper bounds of each box indicate the 25th and 75 percentiles, respectively, and the thick black line indicates the median. The upper (lower) whisker adds (subtracts) 1.5 times the interquartile difference to the 75th (25th) percentile. Points indicate outliers. The dashed horizontal line indicates the observed rainfall peak.}\label{fig:boxplotCTMAX}
\end{figure}
\begin{figure}[t]
  \noindent\includegraphics[width=\columnwidth,angle=0]{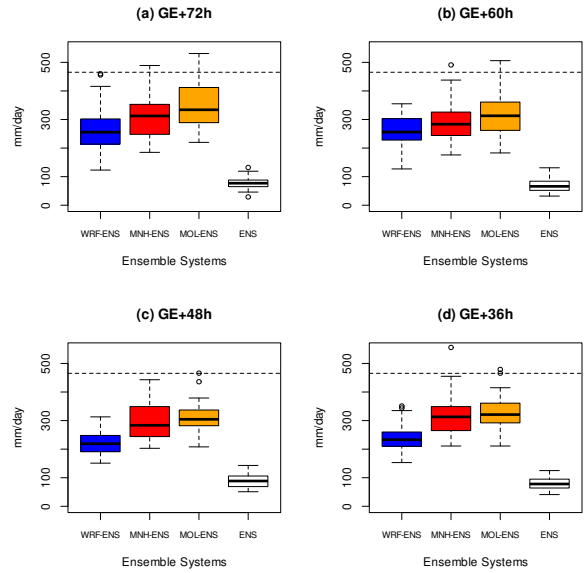}\\
  \caption{As in Figure \ref{fig:boxplotCTMAX} but for the GE case.}\label{fig:boxplotGEMAX}
\end{figure}
\section{Discussions and conclusions}
Reforecasting past extreme weather events is an essential tool to understand the information content of current forecasting systems and monitor the progress achieved in weather modeling, both at the global and regional scale. By using recent versions of the ECMWF IFS model and three regional convection-permitting models, we showed the results obtained for the ensemble reforecasts of the CT and GE heavy-precipitation events that occurred in Italy in autumn 2011. Daily precipitation amounts registered at rain-gauges located in the two areas of interest provided the ground truth to assess the quality of such predictions.

Collective results suggest the potential benefits of running the high-resolution CP ensembles. In fact, using objective verification methods, we demonstrated that CP forecasts outperform ENS predictions for the CT case. This is true in both deterministic (see Figure \ref{fig:PerfDiagCT} and Table \ref{tab:rmse}), and probabilistic terms (see Figure \ref{fig:ROCACT} and Table \ref{tab:ROCA}). As regards the GE case, the results are more controversial. The ENS ensemble mean is more skillful than CP forecasts for the 80-mm threshold (see  panel  (a) in Figure \ref{fig:PerfDiagGE}). We speculate that this happens because the precipitation maxima were observed in a very small portion of the area of interest (namely the Bisagno catchment having an area approximately 100 km$^2$; see \citeauthor{hally2015hydrometeorological} \citeyear{hally2015hydrometeorological}). It is known \citep{gallus2002impact} that in these contexts coarse-resolution models may provide more skillful QPFs than higher-resolution models. For the precipitation threshold equal to 130 mm (see panel (b) in Figure \ref{fig:PerfDiagGE}), any ensemble mean forecast fails to provide useful information. However, looking at the ROC-area profiles for GE (shown in Figure \ref{fig:ROCAGE}), CP ensembles have a better probabilistic precipitation forecast skill than ENS. In fact, the ENS ROC-area profile drops below the critical threshold of 0.5 for precipitation amounts greater than 90-100 mm, whereas CP ensemble ROC-area profiles are  greater than 0.5 for precipitation amounts up to 130-140 mm (which approximately is the median of observed rainfall).
\begin{figure}
  \caption{Output of a single member of the MNH-ENS system: precipitation accumulated every 3 hours (unit of mm) and wind speed and direction averaged over the same period for the CT (panels on the left) and GE (panels on the right) case. The time period (in UTC hours) over which the data are accumulated and averaged is indicated in the top-right corner of each panel. The forecast length is 36 hours. The black point in each panel indicates the location of the rain-gauge that registered the maximum rainfall rate.}\label{fig:MNHconv}
  \includegraphics[width=\columnwidth]{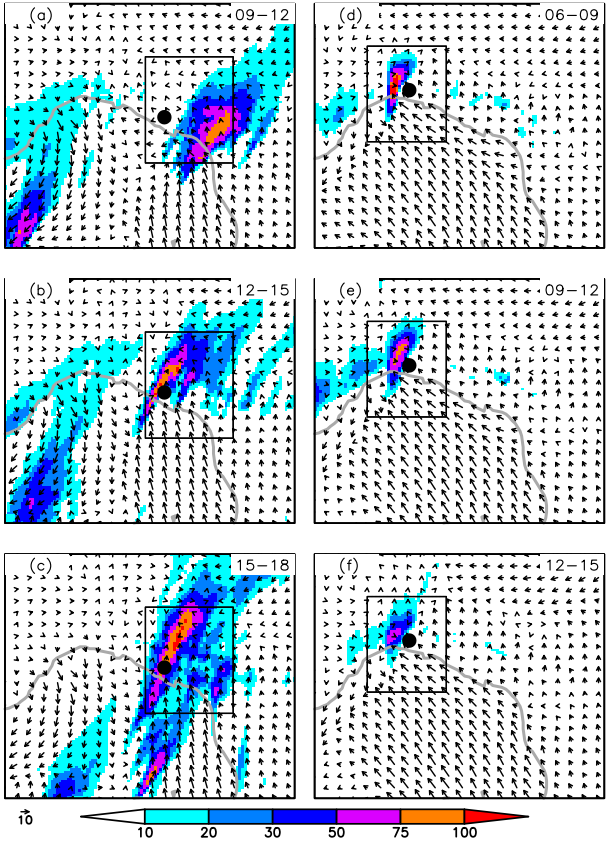}
\end{figure}
As regards the comparison of CP ensemble predictions, the precipitation pattern of the CT case is, in general, best simulated by the MNH-ENS (see Figure \ref{fig:ROCACT} and Table \ref{tab:ROCA}). This could be because the time step used for the simulation is lower than that used for the other two CP models. Indeed, the shorter the time step, the more accurate the prediction \citep{coiffier2011fundamentals}. As regards the GE case, the results produced by the CP ensembles are very similar to each other (see Figure \ref{fig:ROCAGE}), and for some forecast lengths (namely, GE+72h and GE+48h), the WRF-ENS outperforms the other two systems due to small-scale position errors that lead to a reduced number of false alarms. In fact, as Figure \ref{fig:MNHconv} demonstrates, incorrect positioning of the QPF maxima by a few kilometers induces a double-penalty error and impacts the forecast quality assessed by traditional verification statistics. This suggests a few considerations regarding the predictability of the CT and GE cases. Although they share similar synoptic and mesoscale features (see Section \ref{sec:cases}\ref{subsec:description}), as stressed by \cite{davolio2015effects} the CT case is characterized by a greater instability with a level of free convection close to the surface, whereas the GE case exhibits higher levels of convective inhibition, which is overcome by orographic uplift. As a consequence, the rainfall pattern of CT is widespread, whereas that of GE is relatively concentrated along the Ligurian coast and Apenine Mountains. As Figure \ref{fig:PoPGEperc50} shows, ENS data provide only one grid-point with a PoP value higher than 20\%; this leads to an overconfident prediction for the GE case  that none of the CP ensembles are able to mitigate. One may argue that the use of the simple dynamical downscaling method is not suitable to initialize CP ensembles. In fact, the uncertainties in the small-scale features, that are not captured by the large-scale models can lead to the rapid growth of the errors such that the predictability is strongly limited \citep{hohenegger2007predictability}. However, the contamination of the small-scale uncertainty on the whole integration domain depends on the synoptic situation and it is overwhelmed by the influence of lateral boundary conditions when strong synoptic forcings are met (as in the GE case, \citeauthor{rebora2013extreme} \citeyear{rebora2013extreme}). Looking ahead in the near future, when higher-resolution global data assimilation tecniques will produce more accurate analyses, the uncertainty, even at the meso-$\alpha$ and meso-$\beta$ scales, can be better addressed by sampling the members of the global ensemble. Our approach is also justified by recently published papers. For instance, \cite{schwartz2019medium} investigated the value of a CP ensemble directly nested into the NCEP's operational global ensemble data and found that the 3-km ensemble outperforms coarser-resolution ensembles. The author tested his system in the ``extended short-range'', that is, for lead times longer than 24 hours and shorter than 120 hours. Consequently, we deduce that our approach (dynamical downscaling of high-resolution global ensembles) is reasonable in this time range. On the other hand, for forecast lengths shorter than 36 or 24 hours, the design of any CP ensemble should adopt a strategy to perturb the initial conditions to account for small-scale uncertainties.
\begin{figure}[t]
  \noindent\includegraphics[width=\columnwidth,angle=0]{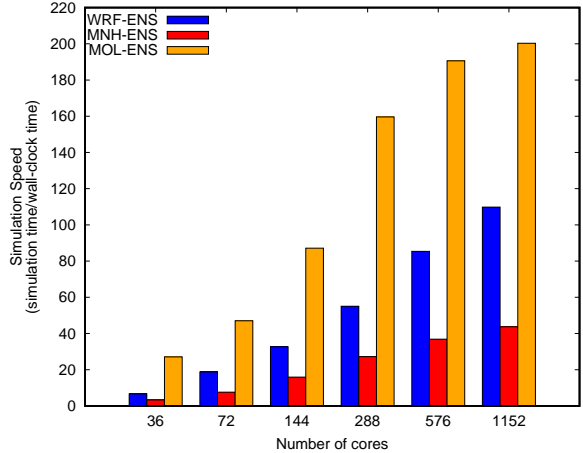}\\
  \caption{Simulation speed, defined as the simulation length (36 hours in the present time) over the elapsed wall-clock time, as a function of the number of cores. For each model, the simulation speed is the average of five selected members of the CT$+$36h simulation.}\label{fig:runtime}
\end{figure}
Figure \ref{fig:boxplotCTMAX} shows that all the members of the four EPS predict QPF maxima that largely underestimate the maximum rainfall amount observed during the CT case. This confirms the findings of similar studies \citep{buzzi2014heavy,davolio2015effects,capecchi2015statistical}. However, we note that some members provide QPF maxima close to  400 mm (see, for instance, WRF-ENS and MOL-ENS in panel (b)), comparable to the outputs of deterministic forecasts run at a much higher horizontal resolution ($\simeq$ 1 km grid spacing, see  \citeauthor{cassola2015numerical} \citeyear{cassola2015numerical}). This confirms \citep{schwartz2019medium} that for forecast lengths in the extended short-range (e.g., from +36 hours to +72 hours), it is worth running probabilistic predictions with a grid spacing of approximately 2.5-3 km to achieve results similar to higher-resolution forecasts in the short-range (forecast lengths less than 24 hours). Figure \ref{fig:boxplotGEMAX} demonstrates that some members of the CP ensembles (see for instance MNH-ENS and MOL-ENS in panel (d)) provide QPF maxima that approach or exceed the maximum observed rainfall value during the GE case. This is consistent with what is shown in Figure \ref{fig:MNHconv} (panels (d), (e) and (f)), which confirms \citep{buzzi2014heavy} that correctly predicting the position of the convergence line over the Ligurian Sea is crucial for generating a rainfall band carrying a large amount of precipitation over the same area.

\begin{figure}[t]
  \noindent\includegraphics[width=\columnwidth,angle=0]{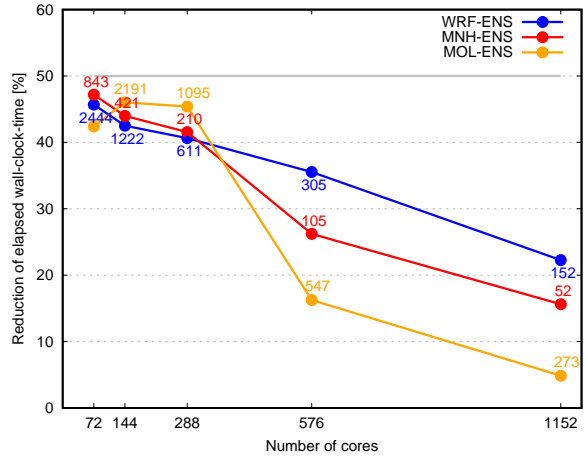}\\
  \caption{Reduction (in percent) of the elapsed wall-clock time when doubling the number of computing cores. In a strong scaling regime, the colored lines should approximate the 50\% horizontal line. The colored labels indicate the number of horizontal grid points assigned to each core.}\label{fig:acc}
\end{figure}

Figure \ref{fig:runtime} shows that the MOLOCH model is the fastest while Meso-NH is the slowest, but Figure \ref{fig:acc} demonstrates that the number of grid points per core $N_{x,y}$ influences the scalability of the three CP models. This is not a new assessment;  our result agrees well with previously published papers regarding the WRF model.  \cite{kruse2013evaluation} found that the WRF model scales approximately linearly through $\simeq$ 650 grid points assigned to each core. Furthermore, the authors concluded that when $N_{x,y}$ is further reduced, the time required to perform the calculations on the perimeter of each patch overwhelms the computational time. For the Meso-NH and MOLOCH models, we found that the gain in elapsed wall-clock time is limited when using 576 cores or more; this occurs when the number of horizontal grid points per core $N_{x,y}$ is less than 105 (547) for the Meso-NH (MOLOCH) model. To the author's knowledge, this an unprecedented assessment regarding these two models. However, we acknowledge that there is no abrupt shift from the strong scaling regime to the weaker regime. The above thresholds can be better defined by smoothly increasing the number of computing elements (instead of doubling it).

\begin{figure}[t]
  \noindent\includegraphics[width=\columnwidth,angle=0]{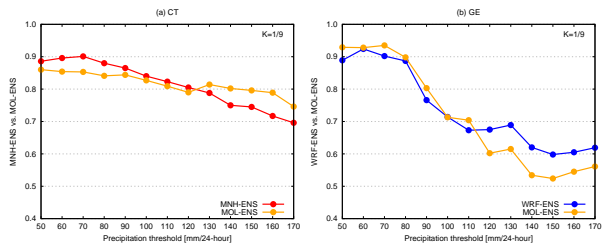}\\
	\caption{Joint evaluation of the performance and simulation speed of the CP ensembles based on the $LISP$ index (the higher the better) defined in equation \ref{eq:lisp}. Panel (a): CT case, MNH-ENS (red line) evaluated against MOL-ENS (orange line). Panel (b): GE case, WRF-ENS (blue line) evaluated against MOL-ENS (orange line).}\label{fig:lisp}
\end{figure}

When investigating different model simulations for the same weather event, it is not straightforward to assess which model provides more reliable information on the basis of both the performance (shown in Figures \ref{fig:ROCACT} and \ref{fig:ROCAGE}) and the computational speed (shown in Figures \ref{fig:runtime} and \ref{fig:acc}). In general, the optimal trade-off between these two terms depends on the end-user requirements. If a large number of cores is available, then the model outputs that, on average, provide the more accurate data should be further analyzed. On the other hand, if the rapid availability of forecasts is crucial for taking the most appropriate action to save lives, properties or for feeding models downstream, then the outputs of the faster model should be examined (provided that its performance is sufficiently fair based on some minimum requirements, i.e., ROC area $>0.5$). In Figure \ref{fig:lisp} we showed the results on the $LISP$ index obtained by setting $\kappa=1/9$, which means that the weights of the performance and speed in equation \ref{eq:lisp} are 0.9 and 0.1, respectively. As regards the CT case (panel (a)), data demonstrate that looking at the MOL-ENS predictions should be preferred for all the precipitation thresholds more than 120 mm. On the other hand, for the precipitation thresholds in the interval [50-120] mm, MNH-ENS provide the more reliable information. As regards the GE case (panel (b)), data show that for precipitation thresholds greater than 120 mm the WRF-ENS ensemble are the more accurate, owing to the better localization of QPF maxima, and they are also the more reliable taking into account the time to realize the predictions. However, we stress the fact that the analysis based on the novel $LISP$ index is strongly influenced by the simulation speed of the MOL-ENS system, which is much higher than that of the other two systems. In fact, the simulation speed of MOL-ENS is, on average, approximately 2.3 and 5.3 times faster than the WRF-ENS and MNH-ENS speeds, respectively. On the other hand, the performances of the three CP ensembles are close to each other, since they amplify the forecasting capability of the global predictions. If we set $\kappa=1/4$, that is the weights of the performance and speed are 0.8 and 0.2 respectively, the MOL-ENS ensemble turns out the more reliable in both cases and for all the precipitation thresholds (maps not shown).

We note that the experimental setup is not the same across the three CP models (see the  settings reported in Table \ref{tab:setup}), which may impact both the model performance and the simulation speed. These experimental setups represent the trade-off between the limited computational resources available and the settings on the horizontal resolution and the extent of the integration domain. To draw meaningful assessments, the horizontal resolution has to be comparable with that of the state-of-the-art regional CP ensembles (see the references cited in Section \ref{sec:intro}) and high enough to partially resolve convective processes. The integration domain has to cover all of Italy to estimate the computational effort needed to deploy a CP ensemble system at the national level. Furthermore, we have to take into account the constraint on the time step, which has to satisfy the numerical stability criterion, and this constraint is not the same across the three CP models. These considerations led to the choices summarized in Table \ref{tab:setup}. We claim that the difference between the WRF grid spacing (3 km) and the Meso-NH and MOLOCH grid spacing (2.5 km) does not remarkably impact the model performance. In fact, as outlined in \cite{buzzi2014heavy}, relevant improvements in the MOLOCH model accuracy are achieved only when the grid spacing is increased to 1.5 km. On the other hand, we underline how the use of a larger domain would be beneficial for the predictions based on the Meso-NH model \citep{davolio2020piedmont}. Because of the small time step needed to guarantee numerical stability, the Meso-NH domain is the smallest one and we speculate that the results presented here most likely underestimate the potential accuracy of the MNH-ENS  forecasts.
Future developments will evaluate the performance of the three CP models in simulating the heavy precipitating event that affected Genoa's town center on the 9th of October 2014. For this event, \cite{fiori2017triggering} concluded that meso-$\gamma$ processes played a crucial role in triggering the convection over the sea in front of the city. Small scale uncertainties grew into upscale uncertainties and contaminated the whole domain \citep{hohenegger2007predictability}. For this case, the simple dynamical downscaling appears inappropriate and different strategies should be adopted to start the CP ensembles.
\datastatement
The ENS data used in this study are accessible through the Meteorological Archival and Retrieval System (MARS) of ECMWF with experiment IDs: gmm2 (CT case), b09s and b09y (GE case).

WRF-ENS, MNH-ENS and MOL-ENS data are available on Zenodo: http://doi.org/10.5281/zenodo.3895551 (CT case), http://doi.org/10.5281/zenodo.3895561 (GE case).

%
 \acknowledgments
The Meso-NH model is freely available under the CeCILL license agreement. 
The author wishes to thank the developers of the model and the User Support for the help.

The MOLOCH model is developed in the framework of the scientific agreement
among ISAC/CNR, ARPAL, ISPRA and LaMMA institutes.

Acknowledgment is made for the use of ECMWF's computing and archive facilities 
within the framework of the SPITCAPE Special Projects during the years 2016-2018 and 2019-2021.

The author would like to express great appreciation to Simon Lang (ECMWF) for his assistance with the IFS model runs 
and to Roberto Buizza (Scuola Superiore Sant'Anna) for his valuable suggestions during the planning of this study.

%
\appendix
\appendixtitle{Model experimental design and impacts on the simulation speed}
The WRF model was compiled on the XC40 Cray supercomputer at ECMWF with  GNU compiler version 4.9.3 (released in 2016) using the distributed and shared memory option. The optimization level was set to ``-O2''.  In this work, we used the WRF-ARW core updated to version 3.7.1 (August 2015).

For the compilation of the Meso-NH model on the XC40 Cray supercomputer at ECMWF, a specific procedure was made available by the model's developers and relies on both Intel and Cray Fortran compilers. For this study, we implemented the version 5.4.2 of the model (released in April 2019) and used Intel compiler version 17.0.3 (released in March 2016) to build the executables. The optimization level was set to ``-O2''.

The ISAC/CNR staff provided full support for the compilation of the MOLOCH model on the ECMWF supercomputer. To the author's knowledge, this was the first time that the model was successfully compiled on a Cray supercomputer architecture. The MOLOCH executables were built in less than 10 minutes with  Intel Fortran compiler version 17.0.3. The more aggressive optimization level ``-O3'' was set. For this work, we implemented the version of the model released in 2018.

Basic information on the model settings and related references are listed in Table A1.
\begin{table*}
  \appendcaption{A1}{References to the radiation, boundary-layer, microphysics and turbulence schemes implemented in the CP models.}\label{tab:app}
  \begin{center}
    \begin{tabular}{p{0.08\linewidth}p{0.25\linewidth}p{0.25\linewidth}p{0.25\linewidth}}
      \hline\hline
\textbf{Scheme} & \textbf{WRF} & \textbf{Meso-NH} & \textbf{MOLOCH}\\
\hline
 \textbf{Radiation}          & Rapid radiative transfer model & ECMWF radiation scheme     & ECMWF radiation scheme\\
                           & \cite{mlawer1997radiative}     & \cite{morcrette2008impact} & \cite{morcrette2008impact}\\\hline
 \textbf{Boundary-layer}     & Yonsei University Scheme & \cite{bougeault1989parameterization} & 1.5-order E-$l$ closure \\
                           & \cite{hong2006new} &  & \cite{zampieri2005sensitivity} \\\hline
 \textbf{Microphysics}       & Thompson Scheme & ICE3 scheme & \cite{drofa2004parameterization}\\
                           & \cite{thompson2008explicit} & \cite{caniaux1994numerical} & \\\hline
 \textbf{Turbulence}         & Yonsei University Scheme $+$ & 2-order
equation & 1.5-order E-$l$ closure\\
        &  2D Smagorisnski  & for turbulent moments   &  \cite{trini2020evaluation}\\
        &  \cite{hong2006new} & \cite{cuxart2000turbulence} & \\
\hline    \end{tabular}
  \end{center}
\end{table*}

The ability to perform detailed ensemble simulations for real-time applications depends on the ability of the code to effectively use large computational resources, which are becoming increasingly more common even at relatively small supercomputing centers. For this reason, we showed the simulation speeds of the WRF, Meso-NH and MOLOCH models in Figure \ref{fig:runtime} and their scaling features (the decrease in the wall-clock time as a function of the number of cores used) in Figure \ref{fig:acc}. The results are intended to provide some guidance to the user interested in having a broad idea about the computational efforts required for accomplishing a CP ensemble simulation. Caution should be taken when looking at the results presented in Figures \ref{fig:runtime} and \ref{fig:acc}. As specified above, different Fortran compilers were used to build the executables, and different optimization levels were set. Recently, newer versions of the WRF executables have become available on the ECMWF supercomputer. These versions are built with the Intel Fortran compiler, which has the reputation of creating faster executables than those produced with the GNU compiler. However, the default compilation Intel Fortran flags of the WRF model are nudged towards accuracy at the expense of performance, with a potential additional decrease in the elapsed wall-clock time up to 5\% when using more aggressive optimization on floating-point data. For these reasons, to have an idea of the performance of the WRF model built with the Intel compiler, we conducted some tests with a more recent version of the WRF model (version 3.9.1, released in August 2017). The preliminary results showed that the WRF executable compiled with the Intel compiler is approximately 30\% faster than that built with the GNU compiler. This conclusion is consistent with the findings in literature \citep{langkamp2011influence,siuta2016viability,moreno2020seeking}. However, no systematic and rigorous sensitivity tests were performed to investigate the impact of fine-tuning compilation flags, hyperthreading or vectorization settings.

In our numerical experiments, the Meso-NH model was the slowest of the three models considered. It must be noted that the adoption of the fourth-order centered advection scheme (CEN4TH), strongly limits the possible values of the time step in satisfying the CFL stability criterion \citep{lunet2017combination}. Some sensitivity tests demonstrated (results not shown) that the time step should not be higher than 6/8 seconds for compliance with the CFL condition. To overcome this constraint, a preliminary test was performed using the less expensive fifth-order weighted essentially nonoscillatory scheme (WENO5), which allows the use of an inflated time step (we set 30 seconds). The results demonstrated a gain in the simulation speed of approximately 18\%. However a systematic verification should be carried out to assess any potential loss in forecast accuracy due to the use of the WENO5 scheme with respect to the CEN4TH scheme.

As regards the simulations performed with the MOLOCH model, we must point out that because of some constraints on the decomposition of the horizontal domain \citep{cioni2014thermal}, the number of grid points is not as constant as the number of cores varies. The larger domain ($514\times650$ grid points) is achieved when using 1152 cores, and the smaller domain ($506\times602$ grid points) is achieved when using 144 cores. Taking into account such differences, the variations in the computational speed of the model are estimated within less than 9\% of the wall-clock time.
%

%
\bibliographystyle{ametsoc2014}
\bibliography{biblio_valcap}

\begin{thebibliography}{79}
\providecommand{\natexlab}[1]{#1}
\providecommand{\url}[1]{\texttt{#1}}
\renewcommand{\UrlFont}{\rmfamily}
\providecommand{\urlprefix}{URL }
\expandafter\ifx\csname urlstyle\endcsname\relax
  \providecommand{\doi}[1]{doi:\discretionary{}{}{}#1}\else
  \providecommand{\doi}{doi:\discretionary{}{}{}\begingroup
  \urlstyle{rm}\Url}\fi
\providecommand{\eprint}[2][]{\url{#2}}

\bibitem[{Berner et~al.(2015)Berner, Fossell, Ha, Hacker,, and
  Snyder}]{berner2015increasing}
Berner, J., K.~Fossell, S.-Y. Ha, J.~Hacker, and C.~Snyder, 2015: Increasing
  the skill of probabilistic forecasts: {U}nderstanding performance
  improvements from model-error representations. \textit{Monthly Weather
  Review}, \textbf{143~(4)}, 1295--1320.

\bibitem[{Bougeault and Lacarrere(1989)Bougeault, and
  Lacarrere}]{bougeault1989parameterization}
Bougeault, P., and P.~Lacarrere, 1989: Parameterization of orography-induced
  turbulence in a mesobeta--scale model. \textit{Monthly Weather Review},
  \textbf{117~(8)}, 1872--1890.

\bibitem[{Bouttier and Raynaud(2018)Bouttier, and
  Raynaud}]{bouttier2018clustering}
Bouttier, F., and L.~Raynaud, 2018: Clustering and selection of boundary
  conditions for limited-area ensemble prediction. \textit{Quarterly Journal of
  the Royal Meteorological Society}, \textbf{144~(717)}, 2381--2391.

\bibitem[{Bouttier et~al.(2016)Bouttier, Raynaud, Nuissier,, and
  M{\'e}n{\'e}trier}]{bouttier2016sensitivity}
Bouttier, F., L.~Raynaud, O.~Nuissier, and B.~M{\'e}n{\'e}trier, 2016:
  Sensitivity of the {AROME} ensemble to initial and surface perturbations
  during {HyMeX}. \textit{Quarterly Journal of the Royal Meteorological
  Society}, \textbf{142}, 390--403.

\bibitem[{Bouttier et~al.(2012)Bouttier, Vi{\'e}, Nuissier,, and
  Raynaud}]{bouttier2012impact}
Bouttier, F., B.~Vi{\'e}, O.~Nuissier, and L.~Raynaud, 2012: Impact of
  stochastic physics in a convection-permitting ensemble. \textit{Monthly
  Weather Review}, \textbf{140~(11)}, 3706--3721.

\bibitem[{Buizza(2019)}]{buizza2019introduction}
Buizza, R., 2019: Introduction to the special issue on ``25 years of ensemble
  forecasting''. \textit{Quarterly Journal of the Royal Meteorological
  Society}, \textbf{145}, 1--11.

\bibitem[{Buizza and Palmer(1995)Buizza, and Palmer}]{buizza1995singular}
Buizza, R., and T.~Palmer, 1995: The singular-vector structure of the
  atmospheric global circulation. \textit{Journal of the Atmospheric Sciences},
  \textbf{52~(9)}, 1434--1456.

\bibitem[{Buzzi et~al.(2014)Buzzi, Davolio, Malguzzi, Drofa,, and
  Mastrangelo}]{buzzi2014heavy}
Buzzi, A., S.~Davolio, P.~Malguzzi, O.~Drofa, and D.~Mastrangelo, 2014: Heavy
  rainfall episodes over {L}iguria in autumn 2011: numerical forecasting
  experiments. \textit{Natural Hazards and Earth System Sciences},
  \textbf{14~(5)}, 1325.

\bibitem[{Caniaux et~al.(1994)Caniaux, Redelsperger,, and
  Lafore}]{caniaux1994numerical}
Caniaux, G., J.~Redelsperger, and J.~P. Lafore, 1994: A numerical study of the
  stratiform region of a fast-moving squall line. {P}art {I}: {G}eneral
  description and water and heat budgets. \textit{Journal of the Atmospheric
  Sciences}, \textbf{51~(14)}, 2046--2074.

\bibitem[{Capecchi et~al.(2015)Capecchi, Perna,, and
  Crisci}]{capecchi2015statistical}
Capecchi, V., M.~Perna, and A.~Crisci, 2015: Statistical modelling of
  rainfall-induced shallow landsliding using static predictors and numerical
  weather predictions: preliminary results. \textit{Natural Hazards and Earth
  System Science}, \textbf{15~(1)}, 75--95.

\bibitem[{Cassola et~al.(2015)Cassola, Ferrari,, and
  Mazzino}]{cassola2015numerical}
Cassola, F., F.~Ferrari, and A.~Mazzino, 2015: Numerical simulations of
  {M}editerranean heavy precipitation events with the {WRF} model: {A}
  verification exercise using different approaches. \textit{Atmospheric
  Research}, \textbf{164}, 210--225.

\bibitem[{Cioni(2014)}]{cioni2014thermal}
Cioni, G., 2014: Thermal structure and dynamical modeling of a {M}editerranean
  {T}ropical-like cyclone. Masterthesis, Fisica del Sistema Terra, Universit\'a
  di Bologna.

\bibitem[{Clark(2019)}]{clark2019comparisons}
Clark, A.~J., 2019: Comparisons of {QPF}s derived from single-and multicore
  convection-allowing ensembles. \textit{Weather and Forecasting},
  \textbf{34~(6)}, 1955--1964.

\bibitem[{Coiffier(2011)}]{coiffier2011fundamentals}
Coiffier, J., 2011: \textit{Fundamentals of numerical weather prediction}.
  Cambridge University Press.

\bibitem[{Cuxart et~al.(2000)Cuxart, Bougeault,, and
  Redelsperger}]{cuxart2000turbulence}
Cuxart, J., P.~Bougeault, and J.-L. Redelsperger, 2000: A turbulence scheme
  allowing for mesoscale and large-eddy simulations. \textit{Quarterly Journal
  of the Royal Meteorological Society}, \textbf{126~(562)}, 1--30.

\bibitem[{Davolio et~al.(2020)Davolio, Malguzzi, Drofa, Mastrangelo,, and
  Buzzi}]{davolio2020piedmont}
Davolio, S., P.~Malguzzi, O.~Drofa, D.~Mastrangelo, and A.~Buzzi, 2020: The
  {P}iedmont flood of {N}ovember 1994: a testbed of forecasting capabilities of
  the {CNR-ISAC} meteorological model suite. \textit{Bulletin of Atmospheric
  Science and Technology}, 1--20.

\bibitem[{Davolio et~al.(2015)Davolio, Silvestro,, and
  Malguzzi}]{davolio2015effects}
Davolio, S., F.~Silvestro, and P.~Malguzzi, 2015: Effects of increasing
  horizontal resolution in a convection-permitting model on flood forecasting:
  {T}he 2011 dramatic events in {L}iguria, {I}taly. \textit{Journal of
  Hydrometeorology}, \textbf{16~(4)}, 1843--1856.

\bibitem[{Drofa and Malguzzi(2004)Drofa, and
  Malguzzi}]{drofa2004parameterization}
Drofa, O., and P.~Malguzzi, 2004: Parameterization of microphysical processes
  in a non hydrostatic prediction model. \textit{Proceedings of 14th Intern.
  Conf. on Clouds and Precipitation (ICCP), Bologna, Italy}, 19--23.

\bibitem[{Duc et~al.(2013)Duc, Saito,, and Seko}]{duc2013spatial}
Duc, L., K.~Saito, and H.~Seko, 2013: Spatial-temporal fractions verification
  for high-resolution ensemble forecasts. \textit{Tellus A: Dynamic Meteorology
  and Oceanography}, \textbf{65~(1)}, 18\,171.

\bibitem[{Ebert(2001)}]{ebert2001ability}
Ebert, E.~E., 2001: Ability of a poor man's ensemble to predict the probability
  and distribution of precipitation. \textit{Monthly Weather Review},
  \textbf{129~(10)}, 2461--2480.

\bibitem[{Ebert(2009)}]{ebert2009neighborhood}
Ebert, E.~E., 2009: Neighborhood verification: {A} strategy for rewarding close
  forecasts. \textit{Weather and Forecasting}, \textbf{24~(6)}, 1498--1510.

\bibitem[{Eckel and Mass(2005)Eckel, and Mass}]{eckel2005aspects}
Eckel, F.~A., and C.~F. Mass, 2005: Aspects of effective mesoscale, short-range
  ensemble forecasting. \textit{Weather and Forecasting}, \textbf{20~(3)},
  328--350.

\bibitem[{ECMWF(2016)}]{roadmap2016}
ECMWF, 2016: Strategy 2016-2025, the strength of a common goal.
  \urlprefix\url{https://www.ecmwf.int/sites/default/files/ECMWF_Strategy_2016-2025.pdf},
  1--32 pp.

\bibitem[{Fiori et~al.(2014)Fiori, Comellas, Molini, Rebora, Siccardi, Gochis,
  Tanelli,, and Parodi}]{fiori2014analysis}
Fiori, E., A.~Comellas, L.~Molini, N.~Rebora, F.~Siccardi, D.~Gochis,
  S.~Tanelli, and A.~Parodi, 2014: Analysis and hindcast simulations of an
  extreme rainfall event in the {M}editerranean area: The {G}enoa 2011 case.
  \textit{Atmospheric Research}, \textbf{138}, 13--29.

\bibitem[{Fiori et~al.(2017)Fiori, Ferraris, Molini, Siccardi, Kranzlmueller,,
  and Parodi}]{fiori2017triggering}
Fiori, E., L.~Ferraris, L.~Molini, F.~Siccardi, D.~Kranzlmueller, and
  A.~Parodi, 2017: Triggering and evolution of a deep convective system in the
  {M}editerranean {S}ea: modelling and observations at a very fine scale.
  \textit{Quarterly Journal of the Royal Meteorological Society},
  \textbf{143~(703)}, 927--941.

\bibitem[{Fresnay et~al.(2012)Fresnay, Hally, Garnaud, Richard,, and
  Lambert}]{fresnay2012heavy}
Fresnay, S., A.~Hally, C.~Garnaud, E.~Richard, and D.~Lambert, 2012: Heavy
  precipitation events in the {M}editerranean: sensitivity to cloud physics
  parameterisation uncertainties. \textit{Natural Hazards \& Earth System
  Sciences}, \textbf{12~(8)}.

\bibitem[{Frogner et~al.(2019)}]{frogner2019harmoneps}
Frogner, I.-L., and Coauthors, 2019: Harmon{EPS}-{T}he {HARMONIE} {E}nsemble
  {P}rediction {S}ystem. \textit{Weather and Forecasting}, \textbf{34~(6)},
  1909--1937.

\bibitem[{Gallus~Jr(2002)}]{gallus2002impact}
Gallus~Jr, W.~A., 2002: Impact of verification grid-box size on warm-season
  {QPF} skill measures. \textit{Weather and forecasting}, \textbf{17~(6)},
  1296--1302.

\bibitem[{Gasperoni et~al.(2020)Gasperoni, Wang,, and
  Wang}]{gasperoni2020comparison}
Gasperoni, N.~A., X.~Wang, and Y.~Wang, 2020: A comparison of methods to sample
  model errors for convection-allowing ensemble forecasts in the setting of
  multiscale initial conditions produced by the {GSI}-based {E}n{V}ar
  assimilation system. \textit{Monthly Weather Review}, \textbf{148~(3)},
  1177--1203.

\bibitem[{Gebhardt et~al.(2008)Gebhardt, Theis, Krahe,, and
  Renner}]{gebhardt2008experimental}
Gebhardt, C., S.~Theis, P.~Krahe, and V.~Renner, 2008: Experimental ensemble
  forecasts of precipitation based on a convection-resolving model.
  \textit{Atmospheric Science Letters}, \textbf{9~(2)}, 67--72.

\bibitem[{Hagelin et~al.(2017)Hagelin, Son, Swinbank, McCabe, Roberts,, and
  Tennant}]{hagelin2017met}
Hagelin, S., J.~Son, R.~Swinbank, A.~McCabe, N.~Roberts, and W.~Tennant, 2017:
  The {M}et {O}ffice convective-scale ensemble, {MOGREPS-UK}. \textit{Quarterly
  Journal of the Royal Meteorological Society}, \textbf{143~(708)}, 2846--2861.

\bibitem[{Hally et~al.(2015)}]{hally2015hydrometeorological}
Hally, A., and Coauthors, 2015: Hydrometeorological multi-model ensemble
  simulations of the 4 {N}ovember 2011 flash flood event in {G}enoa, {I}taly,
  in the framework of the {DRIHM} project. \textit{Natural Hazards and Earth
  System Science}, \textbf{15~(3)}, 537--555.

\bibitem[{Hohenegger and Sch{\"a}r(2007)Hohenegger, and
  Sch{\"a}r}]{hohenegger2007predictability}
Hohenegger, C., and C.~Sch{\"a}r, 2007: Predictability and error growth
  dynamics in cloud-resolving models. \textit{Journal of the atmospheric
  sciences}, \textbf{64~(12)}, 4467--4478.

\bibitem[{Hohenegger et~al.(2008)Hohenegger, Walser, Langhans,, and
  Sch{\"a}r}]{hohenegger2008cloud}
Hohenegger, C., A.~Walser, W.~Langhans, and C.~Sch{\"a}r, 2008: Cloud-resolving
  ensemble simulations of the {A}ugust 2005 {A}lpine flood. \textit{Quarterly
  Journal of the Royal Meteorological Society}, \textbf{134~(633)}, 889--904.

\bibitem[{Hong et~al.(2006)Hong, Noh,, and Dudhia}]{hong2006new}
Hong, S.-Y., Y.~Noh, and J.~Dudhia, 2006: A new vertical diffusion package with
  an explicit treatment of entrainment processes. \textit{Monthly Weather
  Review}, \textbf{134~(9)}, 2318--2341.

\bibitem[{Isaksen et~al.(2010)Isaksen, Bonavita, Buizza, Fisher, Haseler,
  Leutbecher,, and Raynaud}]{isaksen2010ensemble}
Isaksen, L., M.~Bonavita, R.~Buizza, M.~Fisher, J.~Haseler, M.~Leutbecher, and
  L.~Raynaud, 2010: Ensemble of data assimilations at {ECMWF}. Technical
  Memorandum 636, European Centre for Medium-Range Weather Forecasts.

\bibitem[{Jankov et~al.(2019)Jankov, Beck, Wolff, Harrold, Olson, Smirnova,
  Alexander,, and Berner}]{jankov2019stochastically}
Jankov, I., J.~Beck, J.~Wolff, M.~Harrold, J.~B. Olson, T.~Smirnova,
  C.~Alexander, and J.~Berner, 2019: Stochastically perturbed parameterizations
  in an {HRRR}-based ensemble. \textit{Monthly Weather Review},
  \textbf{147~(1)}, 153--173.

\bibitem[{Klasa et~al.(2018)Klasa, Arpagaus, Walser,, and
  Wernli}]{klasa2018evaluation}
Klasa, C., M.~Arpagaus, A.~Walser, and H.~Wernli, 2018: An evaluation of the
  convection-permitting ensemble {COSMO-E} for three contrasting precipitation
  events in {S}witzerland. \textit{Quarterly Journal of the Royal
  Meteorological Society}, \textbf{144~(712)}, 744--764.

\bibitem[{Kruse et~al.(2013)Kruse, Del~Vento, Montuoro, Lubin,, and
  McMillan}]{kruse2013evaluation}
Kruse, C., D.~Del~Vento, R.~Montuoro, M.~Lubin, and S.~McMillan, 2013:
  Evaluation of {WRF} scaling to several thousand cores on the yellowstone
  supercomputer. \textit{Proceedings of the Front Range Consortium for Research
  Computing Conference}.

\bibitem[{K{\"u}hnlein et~al.(2014)K{\"u}hnlein, Keil, Craig,, and
  Gebhardt}]{kuhnlein2014impact}
K{\"u}hnlein, C., C.~Keil, G.~Craig, and C.~Gebhardt, 2014: The impact of
  downscaled initial condition perturbations on convective-scale ensemble
  forecasts of precipitation. \textit{Quarterly Journal of the Royal
  Meteorological Society}, \textbf{140~(682)}, 1552--1562.

\bibitem[{Lac et~al.(2018)}]{lac2018overview}
Lac, C., and Coauthors, 2018: Overview of the {M}eso-{NH} model version 5.4 and
  its applications. \textit{Geoscientific Model Development}, \textbf{11~(5)},
  1929.

\bibitem[{Lang et~al.(2015)Lang, Bonavita,, and Leutbecher}]{lang2015impact}
Lang, S.~T., M.~Bonavita, and M.~Leutbecher, 2015: On the impact of re-centring
  initial conditions for ensemble forecasts. \textit{Quarterly Journal of the
  Royal Meteorological Society}, \textbf{141~(692)}, 2571--2581.

\bibitem[{Langkamp and B\"ohner(2011)Langkamp, and
  B\"ohner}]{langkamp2011influence}
Langkamp, T., and J.~B\"ohner, 2011: Influence of the compiler on multi-{CPU}
  performance of {WRF}v3. \textit{Geoscientific Model Development},
  \textbf{4~(3)}, 611--623.

\bibitem[{Leoncini et~al.(2010)Leoncini, Plant, Gray,, and
  Clark}]{leoncini2010perturbation}
Leoncini, G., R.~S. Plant, S.~L. Gray, and P.~A. Clark, 2010: Perturbation
  growth at the convective scale for {CSIP IOP18}. \textit{Quarterly Journal of
  the Royal Meteorological Society}, \textbf{136~(648)}, 653--670.

\bibitem[{Leutbecher and Palmer(2008)Leutbecher, and
  Palmer}]{leutbecher2008ensemble}
Leutbecher, M., and T.~N. Palmer, 2008: Ensemble forecasting. \textit{Journal
  of Computational Physics}, \textbf{227~(7)}, 3515--3539.

\bibitem[{Loken et~al.(2019)Loken, Clark, Xue,, and Kong}]{loken2019spread}
Loken, E.~D., A.~J. Clark, M.~Xue, and F.~Kong, 2019: Spread and skill in
  mixed-and single-physics convection-allowing ensembles. \textit{Weather and
  Forecasting}, \textbf{34~(2)}, 305--330.

\bibitem[{Lunet et~al.(2017)Lunet, Lac, Auguste, Visentin, Masson,, and
  Escobar}]{lunet2017combination}
Lunet, T., C.~Lac, F.~Auguste, F.~Visentin, V.~Masson, and J.~Escobar, 2017:
  Combination of {WENO} and explicit {R}unge-{K}utta methods for wind transport
  in the {M}eso-{NH} model. \textit{Monthly Weather Review}, \textbf{145~(9)},
  3817--3838.

\bibitem[{Malardel et~al.(2016)Malardel, Wedi, Deconinck, Diamantakis,
  K{\"u}hnlein, Mozdzynski, Hamrud,, and Smolarkiewicz}]{malardel2016new}
Malardel, S., N.~Wedi, W.~Deconinck, M.~Diamantakis, C.~K{\"u}hnlein,
  G.~Mozdzynski, M.~Hamrud, and P.~Smolarkiewicz, 2016: A new grid for the
  {IFS}. \textit{ECMWF Newsl}, \textbf{146}, 23--28.

\bibitem[{Malguzzi et~al.(2006)Malguzzi, Grossi, Buzzi, Ranzi,, and
  Buizza}]{malguzzi20061966}
Malguzzi, P., G.~Grossi, A.~Buzzi, R.~Ranzi, and R.~Buizza, 2006: The 1966
  ``century'' flood in {I}taly: {A} meteorological and hydrological
  revisitation. \textit{Journal of Geophysical Research: Atmospheres},
  \textbf{111~(D24)}.

\bibitem[{Mason(1982)}]{mason1982model}
Mason, I., 1982: A model for assessment of weather forecasts. \textit{Aust.
  Meteor. Mag}, \textbf{30~(4)}, 291--303.

\bibitem[{Miglietta and Rotunno(2009)Miglietta, and
  Rotunno}]{miglietta2009numerical}
Miglietta, M.~M., and R.~Rotunno, 2009: Numerical simulations of conditionally
  unstable flows over a mountain ridge. \textit{Journal of the atmospheric
  sciences}, \textbf{66~(7)}, 1865--1885.

\bibitem[{Mlawer et~al.(1997)Mlawer, Taubman, Brown, Iacono,, and
  Clough}]{mlawer1997radiative}
Mlawer, E.~J., S.~J. Taubman, P.~D. Brown, M.~J. Iacono, and S.~A. Clough,
  1997: Radiative transfer for inhomogeneous atmospheres: {RRTM}, a validated
  correlated-k model for the longwave. \textit{Journal of Geophysical Research:
  Atmospheres}, \textbf{102~(D14)}, 16\,663--16\,682.

\bibitem[{Molteni et~al.(1996)Molteni, Buizza, Palmer,, and
  Petroliagis}]{molteni1996ecmwf}
Molteni, F., R.~Buizza, T.~N. Palmer, and T.~Petroliagis, 1996: The {ECMWF}
  ensemble prediction system: Methodology and validation. \textit{Quarterly
  journal of the royal meteorological society}, \textbf{122~(529)}, 73--119.

\bibitem[{Montani et~al.(2011)Montani, Cesari, Marsigli,, and
  Paccagnella}]{montani2011seven}
Montani, A., D.~Cesari, C.~Marsigli, and T.~Paccagnella, 2011: Seven years of
  activity in the field of mesoscale ensemble forecasting by the {COSMO-LEPS}
  system: main achievements and open challenges. \textit{Tellus A: Dynamic
  Meteorology and Oceanography}, \textbf{63~(3)}, 605--624.

\bibitem[{Morcrette et~al.(2008)Morcrette, Barker, Cole, Iacono,, and
  Pincus}]{morcrette2008impact}
Morcrette, J., H.~W. Barker, J.~Cole, M.~J. Iacono, and R.~Pincus, 2008: Impact
  of a new radiation package, {M}c{R}ad, in the {ECMWF} {I}ntegrated
  {F}orecasting {S}ystem. \textit{Monthly Weather Review}, \textbf{136~(12)},
  4773--4798.

\bibitem[{Moreno et~al.(2020)Moreno, Arias, Cazorla, Pardo,, and
  Tapiador}]{moreno2020seeking}
Moreno, R., E.~Arias, D.~Cazorla, J.~Pardo, and F.~Tapiador, 2020: Seeking the
  best {W}eather {R}esearch and {F}orecasting model performance: an empirical
  score approach. \textit{The Journal of Supercomputing}, \textbf{~(76)},
  9629--9653.

\bibitem[{Peralta et~al.(2012)Peralta, Ben~Bouall{\`e}gue, Theis, Gebhardt,,
  and Buchhold}]{peralta2012accounting}
Peralta, C., Z.~Ben~Bouall{\`e}gue, S.~Theis, C.~Gebhardt, and M.~Buchhold,
  2012: Accounting for initial condition uncertainties in {COSMO-DE-EPS}.
  \textit{Journal of Geophysical Research: Atmospheres}, \textbf{117~(D7)}.

\bibitem[{Raynaud and Bouttier(2016)Raynaud, and
  Bouttier}]{raynaud2015comparison}
Raynaud, L., and F.~Bouttier, 2016: Comparison of initial perturbation methods
  for ensemble prediction at convective scale. \textit{Quarterly Journal of the
  Royal Meteorological Society}, \textbf{142~(695)}, 854--866.

\bibitem[{Raynaud and Bouttier(2017)Raynaud, and Bouttier}]{raynaud2017impact}
Raynaud, L., and F.~Bouttier, 2017: The impact of horizontal resolution and
  ensemble size for convective-scale probabilistic forecasts. \textit{Quarterly
  Journal of the Royal Meteorological Society}, \textbf{143~(709)}, 3037--3047.

\bibitem[{Rebora et~al.(2013)}]{rebora2013extreme}
Rebora, N., and Coauthors, 2013: Extreme rainfall in the {M}editerranean: what
  can we learn from observations? \textit{Journal of Hydrometeorology},
  \textbf{14~(3)}, 906--922.

\bibitem[{Roebber(2009)}]{roebber2009visualizing}
Roebber, P.~J., 2009: Visualizing multiple measures of forecast quality.
  \textit{Weather and Forecasting}, \textbf{24~(2)}, 601--608.

\bibitem[{Romine et~al.(2014)Romine, Schwartz, Berner, Fossell, Snyder,
  Anderson,, and Weisman}]{romine2014representing}
Romine, G.~S., C.~S. Schwartz, J.~Berner, K.~R. Fossell, C.~Snyder, J.~L.
  Anderson, and M.~L. Weisman, 2014: Representing forecast error in a
  convection-permitting ensemble system. \textit{Monthly Weather Review},
  \textbf{142~(12)}, 4519--4541.

\bibitem[{Rotunno and Ferretti(2001)Rotunno, and
  Ferretti}]{rotunno2001mechanisms}
Rotunno, R., and R.~Ferretti, 2001: Mechanisms of intense {A}lpine rainfall.
  \textit{Journal of the Atmospheric Sciences}, \textbf{58~(13)}, 1732--1749.

\bibitem[{Schwartz(2019)}]{schwartz2019medium}
Schwartz, C.~S., 2019: Medium-range convection-allowing ensemble forecasts with
  a variable-resolution global model. \textit{Monthly Weather Review},
  \textbf{147~(8)}, 2997--3023.

\bibitem[{Schwartz et~al.(2017)Schwartz, Romine, Fossell, Sobash,, and
  Weisman}]{schwartz2017toward}
Schwartz, C.~S., G.~S. Romine, K.~R. Fossell, R.~A. Sobash, and M.~L. Weisman,
  2017: Toward 1-km ensemble forecasts over large domains. \textit{Monthly
  Weather Review}, \textbf{145~(8)}, 2943--2969.

\bibitem[{Schwartz et~al.(2014)Schwartz, Romine, Smith,, and
  Weisman}]{schwartz2014characterizing}
Schwartz, C.~S., G.~S. Romine, K.~R. Smith, and M.~L. Weisman, 2014:
  Characterizing and optimizing precipitation forecasts from a
  convection-permitting ensemble initialized by a mesoscale ensemble {K}alman
  filter. \textit{Weather and Forecasting}, \textbf{29~(6)}, 1295--1318.

\bibitem[{Schwartz et~al.(2015{\natexlab{a}})Schwartz, Romine, Sobash,
  Fossell,, and Weisman}]{schwartz2015ncar}
Schwartz, C.~S., G.~S. Romine, R.~A. Sobash, K.~R. Fossell, and M.~L. Weisman,
  2015{\natexlab{a}}: {NCAR}{\rq}s experimental real-time convection-allowing
  ensemble prediction system. \textit{Weather and Forecasting},
  \textbf{30~(6)}, 1645--1654.

\bibitem[{Schwartz et~al.(2015{\natexlab{b}})Schwartz, Romine, Weisman, Sobash,
  Fossell, Manning,, and Trier}]{schwartz2015real}
Schwartz, C.~S., G.~S. Romine, M.~L. Weisman, R.~A. Sobash, K.~R. Fossell,
  K.~W. Manning, and S.~B. Trier, 2015{\natexlab{b}}: A real-time
  convection-allowing ensemble prediction system initialized by mesoscale
  ensemble {K}alman filter analyses. \textit{Weather and Forecasting},
  \textbf{30~(5)}, 1158--1181.

\bibitem[{Siuta et~al.(2016)Siuta, West, Modzelewski, Schigas,, and
  Stull}]{siuta2016viability}
Siuta, D., G.~West, H.~Modzelewski, R.~Schigas, and R.~Stull, 2016: Viability
  of cloud computing for real-time numerical weather prediction.
  \textit{Weather and Forecasting}, \textbf{31~(6)}, 1985--1996.

\bibitem[{Skamarock et~al.(2008)}]{skamarock2008description}
Skamarock, W.~C., and Coauthors, 2008: A description of the {A}dvanced
  {R}esearch {WRF} version 3. \textit{NCAR Tech. Note NCAR/TN-475+ STR},
  \doi{10.5065/D68S4MVH}.

\bibitem[{Thompson et~al.(2008)Thompson, Field, Rasmussen,, and
  Hall}]{thompson2008explicit}
Thompson, G., P.~R. Field, R.~M. Rasmussen, and W.~D. Hall, 2008: Explicit
  forecasts of winter precipitation using an improved bulk microphysics scheme.
  {P}art {II}: Implementation of a new snow parameterization. \textit{Monthly
  Weather Review}, \textbf{136~(12)}, 5095--5115.

\bibitem[{Tiesi et~al.(2016)Tiesi, Miglietta, Conte, Drofa, Davolio, Malguzzi,,
  and Buzzi}]{tiesi2016heavy}
Tiesi, A., M.~M. Miglietta, D.~Conte, O.~Drofa, S.~Davolio, P.~Malguzzi, and
  A.~Buzzi, 2016: Heavy rain forecasting by model initialization with {LAPS}: a
  case study. \textit{IEEE Journal of Selected Topics in Applied Earth
  Observations and Remote Sensing}, \textbf{9~(6)}, 2619--2627.

\bibitem[{Toth and Kalnay(1993)Toth, and Kalnay}]{toth1993ensemble}
Toth, Z., and E.~Kalnay, 1993: Ensemble forecasting at {NMC}: {T}he generation
  of perturbations. \textit{Bulletin of the american meteorological society},
  \textbf{74~(12)}, 2317--2330.

\bibitem[{Tracton and Kalnay(1993)Tracton, and Kalnay}]{tracton1993operational}
Tracton, M.~S., and E.~Kalnay, 1993: Operational ensemble prediction at the
  {N}ational {M}eteorological {C}enter: {P}ractical aspects. \textit{Weather
  and Forecasting}, \textbf{8~(3)}, 379--398.

\bibitem[{Trini~Castelli et~al.(2020)Trini~Castelli, Bisignano, Donateo, Landi,
  Martano,, and Malguzzi}]{trini2020evaluation}
Trini~Castelli, S., A.~Bisignano, A.~Donateo, T.~C. Landi, P.~Martano, and
  P.~Malguzzi, 2020: Evaluation of the turbulence parametrization in the
  {MOLOCH} meteorological model. \textit{Quarterly Journal of the Royal
  Meteorological Society}, \textbf{146~(726)}, 124--140.

\bibitem[{Vi{\'e} et~al.(2011)Vi{\'e}, Nuissier,, and Ducrocq}]{vie2011cloud}
Vi{\'e}, B., O.~Nuissier, and V.~Ducrocq, 2011: Cloud-resolving ensemble
  simulations of {M}editerranean heavy precipitating events: uncertainty on
  initial conditions and lateral boundary conditions. \textit{Monthly Weather
  Review}, \textbf{139~(2)}, 403--423.

\bibitem[{Wang et~al.(2011)}]{wang2011central}
Wang, Y., and Coauthors, 2011: The {C}entral {E}uropean limited-area ensemble
  forecasting system: {ALADIN-LAEF}. \textit{Quarterly Journal of the Royal
  Meteorological Society}, \textbf{137~(655)}, 483--502.

\bibitem[{Wilks(2011)}]{wilks2011statistical}
Wilks, D.~S., 2011: \textit{Statistical methods in the atmospheric sciences},
  Vol. 100. Academic press.

\bibitem[{Zampieri et~al.(2005)Zampieri, Malguzzi,, and
  Buzzi}]{zampieri2005sensitivity}
Zampieri, M., P.~Malguzzi, and A.~Buzzi, 2005: Sensitivity of quantitative
  precipitation forecasts to boundary layer parameterization: a flash flood
  case study in the {W}estern {M}editerranean. \textit{Natural Hazards and
  Earth System Science}, \textbf{5~(4)}, 603--612.

\end{thebibliography}


%

%
\end{document}